\documentclass{aa}

\usepackage[margin=0.7in]{geometry}

\usepackage{graphicx}
\usepackage{wrapfig}
\usepackage{lscape}
\usepackage{color}

\newcommand{\beq}{\begin{equation}}
\newcommand{\eeq}{\end{equation}}
\newcommand{\beqa}{\begin{eqnarray}}
\newcommand{\eeqa}{\end{eqnarray}}

\begin{document}

\title{Numerical simulations of spicule formation in the solar atmosphere}

\author{K. Murawski\inst{1} \and T.V. Zaqarashvili\inst{2,3}}

 \institute{Group of Astrophysics,
             UMCS, ul. Radziszewskiego 10, 20-031 Lublin, Poland\\
              \email{kmur@kft.umcs.lublin.pl}
               \and
            Space Research Institute, Austrian Academy of Sciences, Schmiedlstrasse 6, 8042 Graz, Austria\\
             \email{teimuraz.zaqarashvili@oeaw.ac.at}
               \and
            Abastumani Astrophysical Observatory at Ilia State University, Kazbegi ave. 2a, Tbilisi, Georgia
                  }

\date{Received / Accepted }

\abstract
{We study the upward propagation of a localized velocity pulse that is initially launched below the transition region within the solar atmosphere.
The pulse quickly steepens into a shock, which may lead to the formation of spicules.
}
{We aim to explore the spicule formation scenario in the framework of the rebound shock model.
}
{We solve two-dimensional time-dependent magnetohydrodynamic equations numerically to find spatial and
temporal dynamics of spicules.
}
{The numerical simulations show that the strong initial pulse may lead to the quasi periodic rising of
chromospheric material into the lower corona in the form of spicules.
The periodicity
results from
the nonlinear wake that is formed behind the pulse in the stratified atmosphere.
The superposition of raising and falling off plasma portions resembles the time sequence of single and
double (sometimes even triple) spicules, which is consistent with observational findings.
}
{The two-dimensional rebound shock model may explain the observed speed, width, and heights of type I spicules, as well
as observed multi-structural and bi-directional flows. The model also predicts the appearance of spicules with
$3-5$ min period due to the consecutive shocks.

}


\titlerunning{Spicule formation in the solar atmosphere}

\authorrunning{}

\maketitle
\section{Introduction}
Although spicules were discovered almost 130 years ago (Secchi \cite{secchi1877}),
they still remain one of mysterious
phenomena
in the solar atmosphere. Spicules are usually detected in chromospheric H$_\alpha$, D$_3$, and Ca II H lines as thin
and elongated structures in the solar limb.
Many authors consider H$_\alpha$ dark mottles as a disc counterpart of limb spicules
(Suematsu et al. \cite{Suematsu1995}), however this point is still under discussion.
Excellent reviews
of the general properties of spicules (and mottles)
were presented almost forty years ago by Beckers (\cite{bec68,bec72}). Suematsu (\cite{Suematsu1998}) and
Sterling (\cite{Sterling2000}) published more recent reviews of
observational and theoretical aspects of spicules, respectively,
while Zaqarashvili \& Erd{\'e}lyi (\cite{Zaqarashvili2009})
summarized observed oscillation events.
Very recently, Pasachoff et al. (\cite{pas09}) studied high-resolution dynamics of limb spicules
by Swedish Solar Telescope (SST) with 0.2 arc sec resolution. High resolution observations by
Solar Optical Telescope (SOT) on board of the recently launched Hinode show another type of spicules,
which have quite different properties than classical limb spicules (De Pontieu et al. \cite{dep07b}).
The classical and newly observed spicules are referred to as types I and II, respectively.

Disc counterparts of limb spicules still are not known despite of all the discussion over the years
(Tsiropoula et al. \cite{Tsiropoula1994}, Sterling \cite{Sterling2000}).
However, recent high-resolution observations on SST and Hinode/SOT suggest that active-region dynamic fibrils
(Hansteen et al. \cite{hansten2006}, De Pontieu et al. \cite{dep07}) and quiet Sun mottles (Rouppe van der Voort et al. \cite{rouppe2007})
have properties similar to type I spicules. On the other hand, Rouppe van der Voort et al. (\cite{rouppe2009})
report discovering of disc counterparts to type II spicules through observations on SST.

Despite all the mechanisms of spicule formation
proposed in the literature
none of them provides convincing explanations of all properties.
Spicule formation mechanisms can be formally divided into three different groups:
pulses, Alfv\'en waves, and p-mode leakage.

The main idea of spicule formation
by impulsively launched perturbations
is as follows. Velocity or gas pressure pulses deposited in the lower atmosphere
are steepened into shocks as a result of the rapid decrease in mass density with height.
These shocks lift up the transition region, producing spicules.
Hollweg (\cite{hol82}) developed the shock model with one-dimensional simulations.
In his rebound shock model, Hollweg simulated gas pressure to lift the transition region to the observed altitudes.
Suematsu et al. (\cite{Suematsu1982}) did the same for velocity pulses. Later on,
Sterling et al. (\cite{Sterling1990,Sterling1993}), Cheng (\cite{cheng1992}), and
Heggland et al. (\cite{heg07}) studied the shock models, including radiation and heat conduction.

Hollweg et al. (\cite{holw82}) showed that Alfv\'en waves may be nonlinearly coupled to fast magnetoacoustic shocks,
which may lift up the transition region. Later on, Kudoh \& Shibata (\cite{kud99}) suggested that the random nonlinear
Alfv\'enic pulses may reproduce the spicules at the observed heights.
On the other hand, the damping of high-frequency Alfv\'en waves due to ion-neutral collisions
may be responsible for formation of spicules (Haerendel \cite{Haerendel1992}, James et al. \cite{james03}).

De Pontieu et al. (2004) propose that, as a result of fall off of acoustic cut-off frequency,
p-modes may be channeled into the solar corona along inclined magnetic field lines.
The oscillations then may be steepened into shocks producing spicules. De Pontieu et al. (2004)
argue that the observed quasi $5$-min period in spicule appearance is associated with the periodicity of p-modes.
Several papers (Hansteen et al. \cite{hansten2006}, De Pontieu et al. \cite{dep07}) then studied the formation
of spicules due to the leakage of photospheric convective motions and oscillations into upper atmosphere.
By performing numerical simulations of MHD equations the authors show that
oscillations and flows generated by self consistent two-dimensional (2D) convective motions drive spicules.

Numerous other mechanisms of spicule formation were suggested, such as
resonant buffeting of anchored magnetic flux tubes by solar granules (Roberts \cite{roberts1979}).
Detailed summary of these mechanisms can be found in the review paper by Sterling (\cite{Sterling2000}).
Horizontal magnetic flux emergence has also been found to be one of many
possible spicule drivers, the others being convective overshoot,
collapse of granules, p-modes, and reconnection in either the
photosphere or chromosphere (Mart\'inez-Sykora et al. \cite{martinez09}).

Despite significant achievements in the above-mentioned papers, there are still no existing mechanisms
that explain the observed double structures of spicules (Tanaka \cite{Tanaka1974}, Dara et al. \cite{Dara1998},
Suematsu et al. \cite{Suematsu2008}) and bi-directional flow (Tsiropoula et al. \cite{Tsiropoula1994},
Tziotziou et al. \cite{Tziotziou2003,Tziotziou2004}, Pasachoff et al. \cite{pas09}).
Suematsu et al. (\cite{Suematsu2008}) and Tsiropoula et al. (\cite{Tsiropoula1994}) have proposed a magnetic reconnection
as an explanation of, respectively, the double structures and bi-directional flow.
However, neither cogent numerical simulations nor analytical models have been presented so far. Therefore, the problem remains still
open to discussion.

In this paper we perform 2D numerical simulations of magnetohydrodynamic (MHD) equations
and show that the 2D rebound shock model may explain both the double structures and bi-directional flow in spicules.
Unlike previous 2D numerical simulations (Hansteen et al. \cite{hansten2006}, De Pontieu et al. \cite{dep07}),
which considered a periodic initial driver, we launch a single initial pulse and follow its temporal development.
Such an approach allows us to trace propagation of the pulse and the internal structure of a spicule in detail,
which explains double structure and bi-directional flow.

This paper is organized as follows. The main properties of limb spicules are summarized in Sect.~2.
A numerical model is presented in Sect.~3, and the corresponding numerical results
are shown in Sect.~4. This paper concludes with a discussion and short summary of the main results in Sects.~5 \& 6, respectively.
\section{Main properties of limb spicules}
A spicule diameter was estimated from ground-based observations to lie
within the range of $700-2500$ km more than four decades ago by Beckers (\cite{bec68}).
Spicules seemed to be wider in the Ca II H line than
in the H$_\alpha$ line (Beckers \cite{bec72}). Pasachoff et al. (\cite{pas09}) estimated the mean diameter of spicules as 660 $\pm$ 200 km
using high-resolution observations of SST in H$_\alpha$, which fits the old observational data.
The type II spicules have smaller diameters ($\le$ 200 km) in the Ca II H line (De Pontieu et al. \cite{dep07b}).

\begin{figure}[h]
\begin{center}
\includegraphics[width=6.1cm,height=7.2cm]{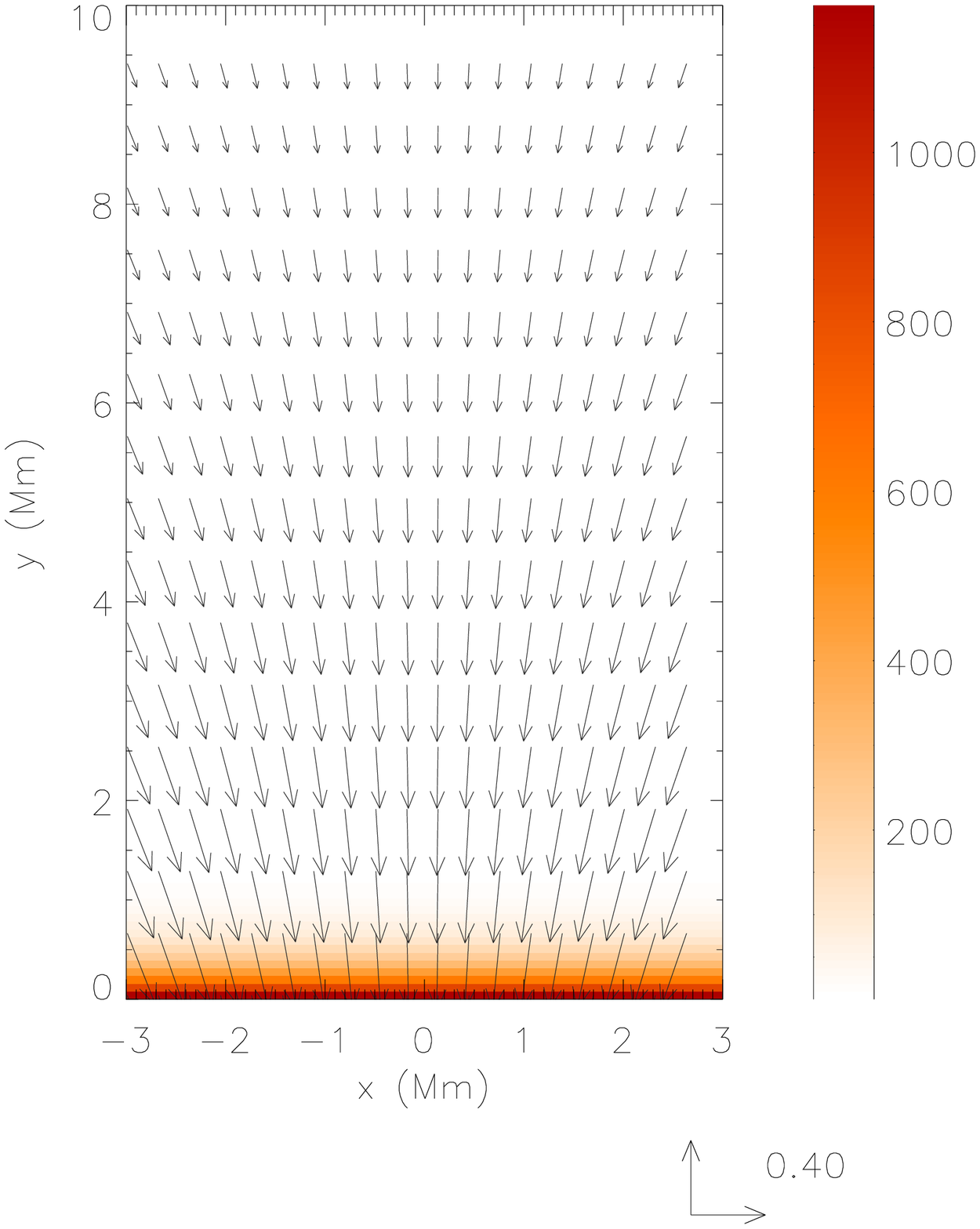}
\includegraphics[width=6.1cm,height=7.2cm]{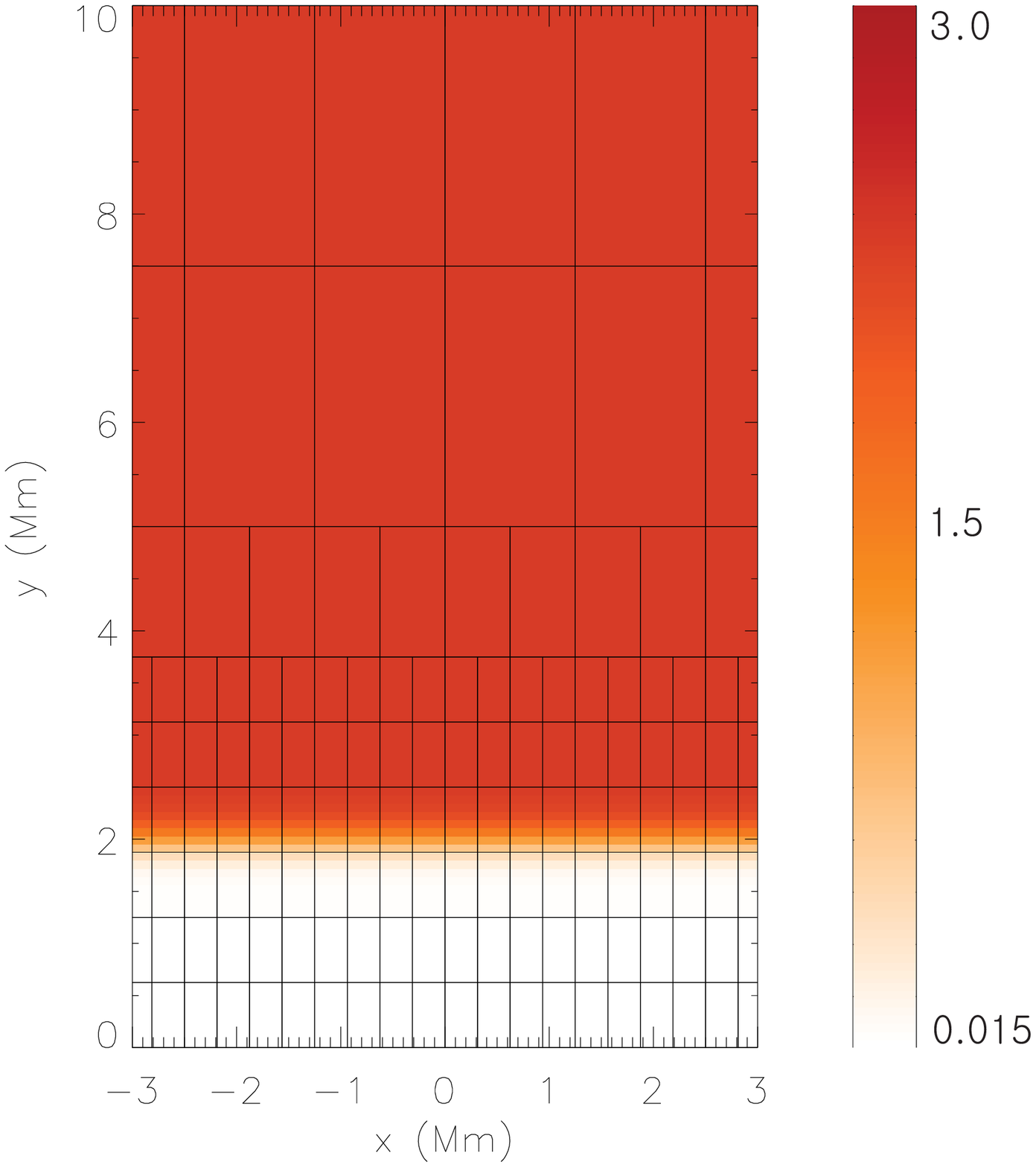}
\caption{\small
Equilibrium configuration of the system: (top panel) magnetic field (in units of
$1.12\times 10^{-3}$ Tesla)
lines are shown as arrows, and the mass density $\varrho_{\rm e}(y)$ (colour map plot)
is expressed in units of $10^{-15}$ kg\,m$^{-3}$; (bottom panel) temperature $T_{\rm e}(y)$
(colour map plot) is drawn in units of $1$ MK  and block boundaries are represented
by solid lines.
}
\label{fig:initial_profile}
\end{center}
\end{figure}
\begin{figure*}
{
\includegraphics[width=6.1cm,height=7.2cm]{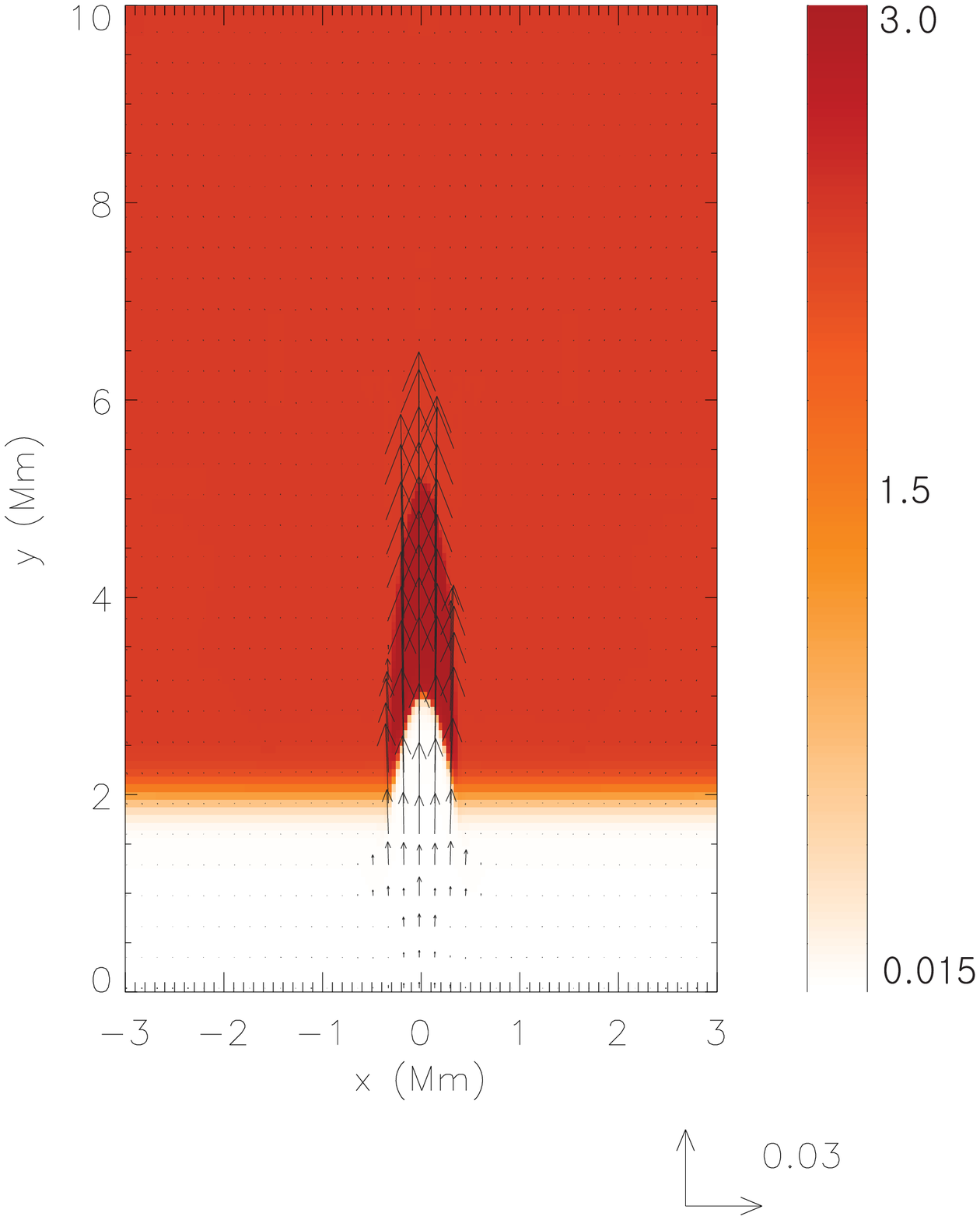}
\includegraphics[width=6.1cm,height=7.2cm]{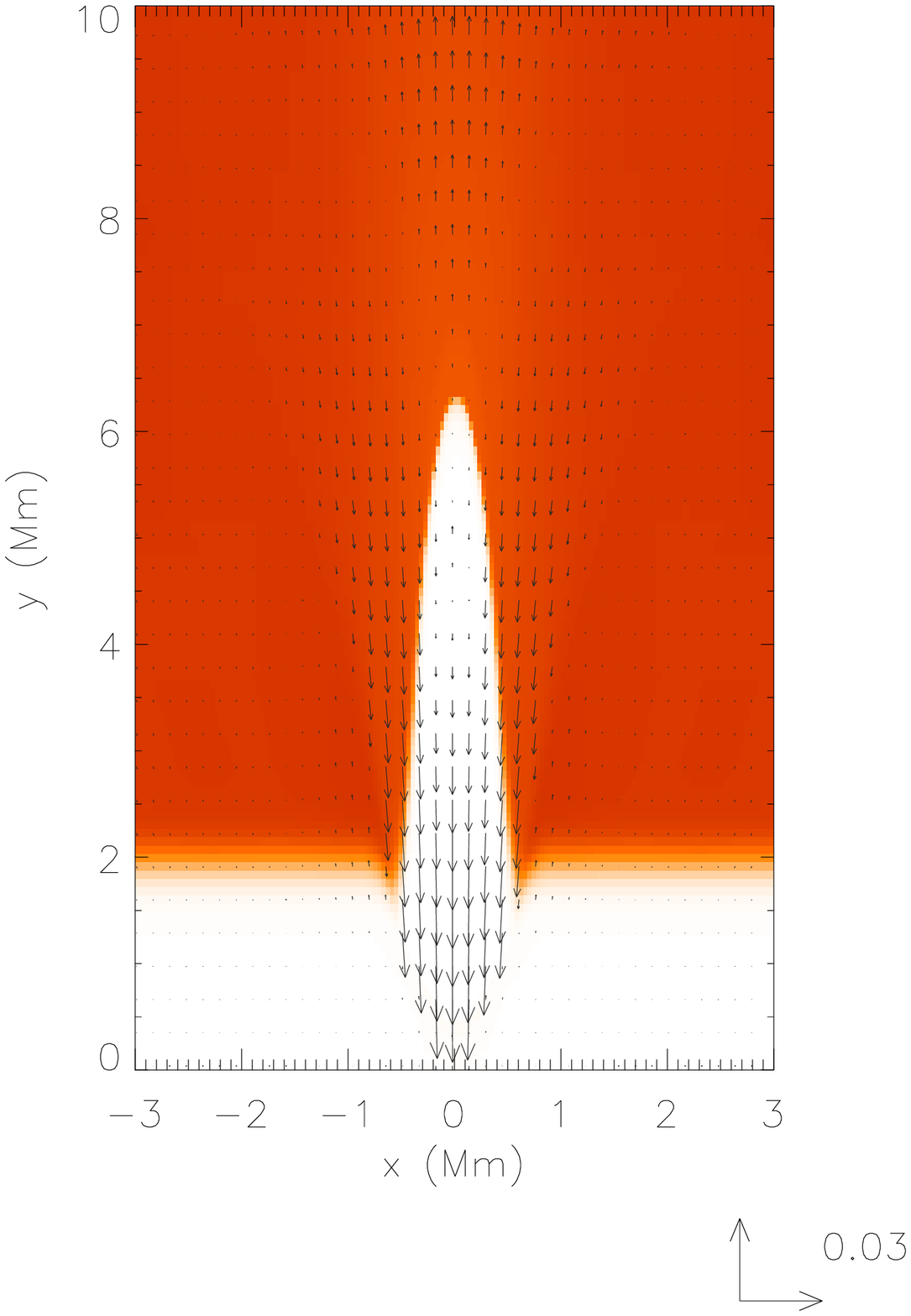}
\includegraphics[width=6.1cm,height=7.2cm]{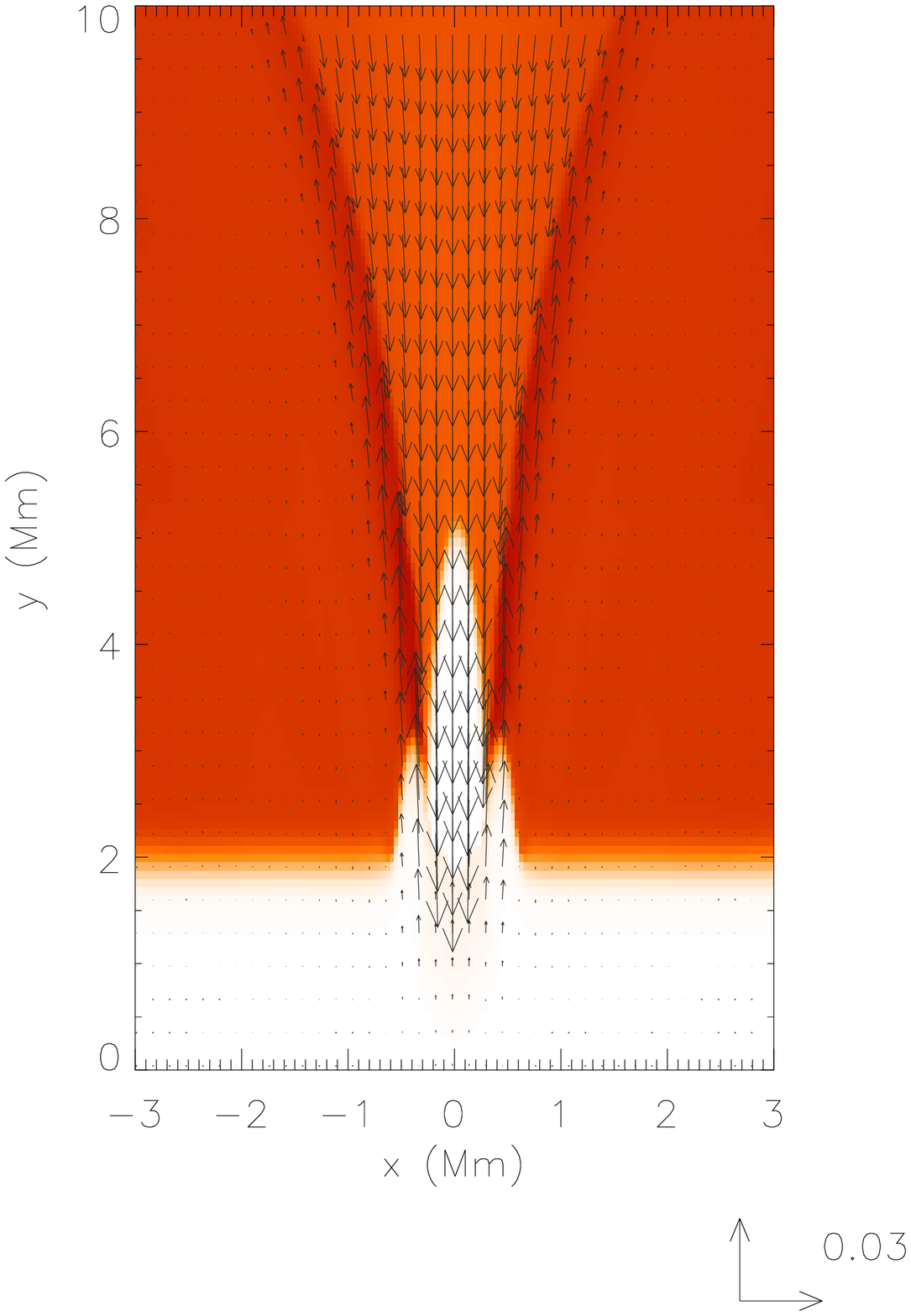}
\includegraphics[width=6.1cm,height=7.2cm]{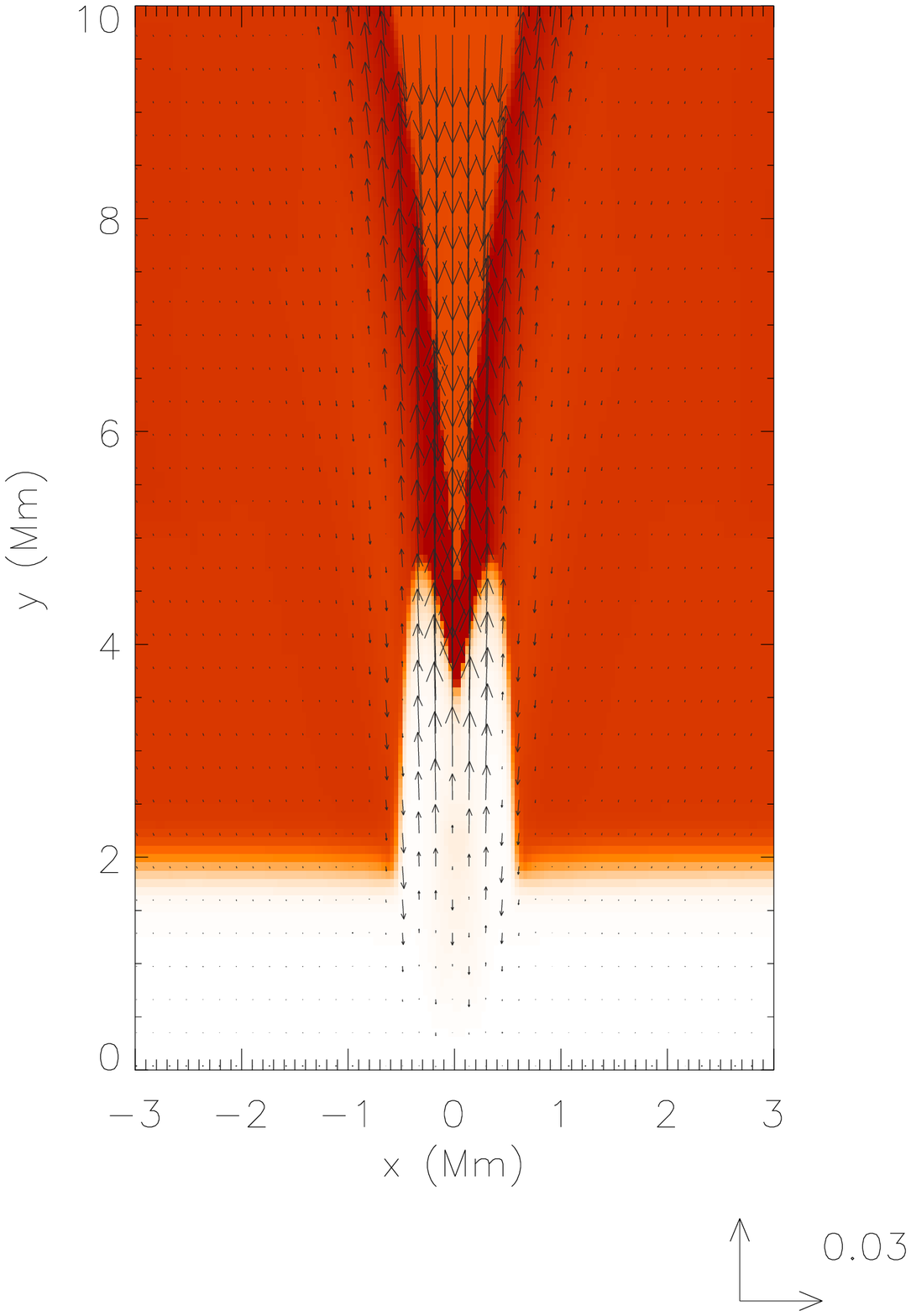}
\includegraphics[width=6.1cm,height=7.2cm]{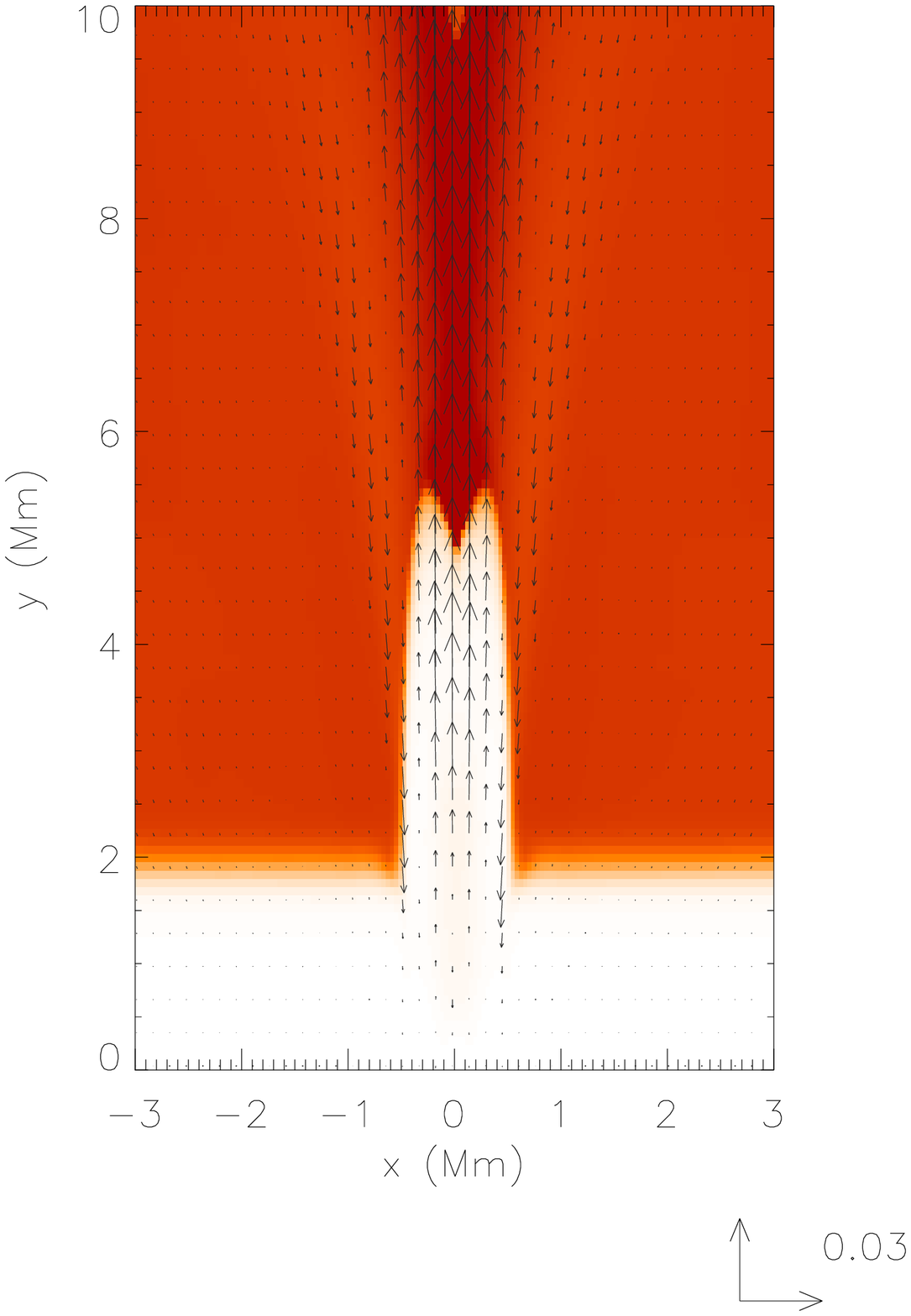}
\includegraphics[width=6.1cm,height=7.2cm]{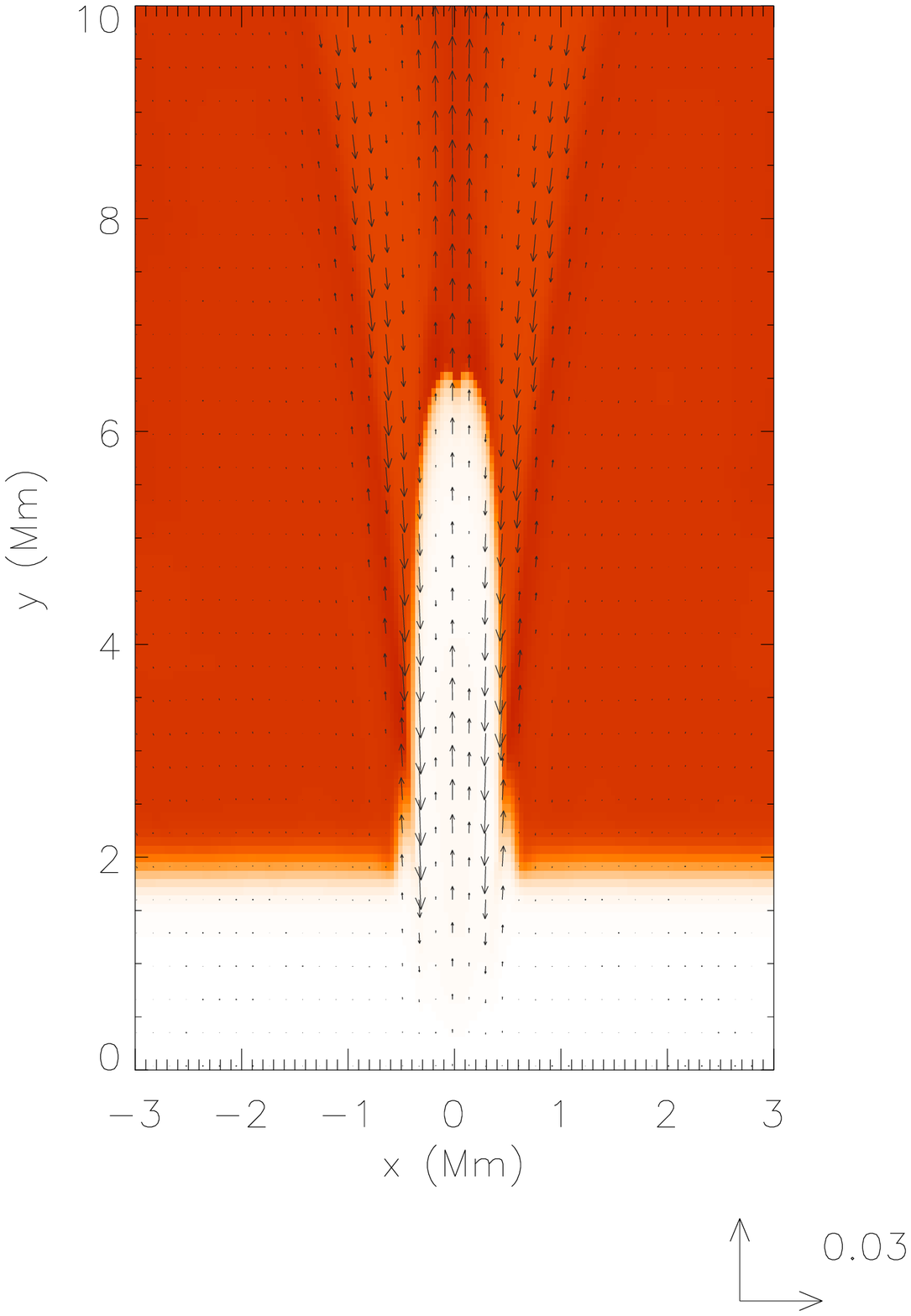}
\includegraphics[width=6.1cm,height=7.2cm]{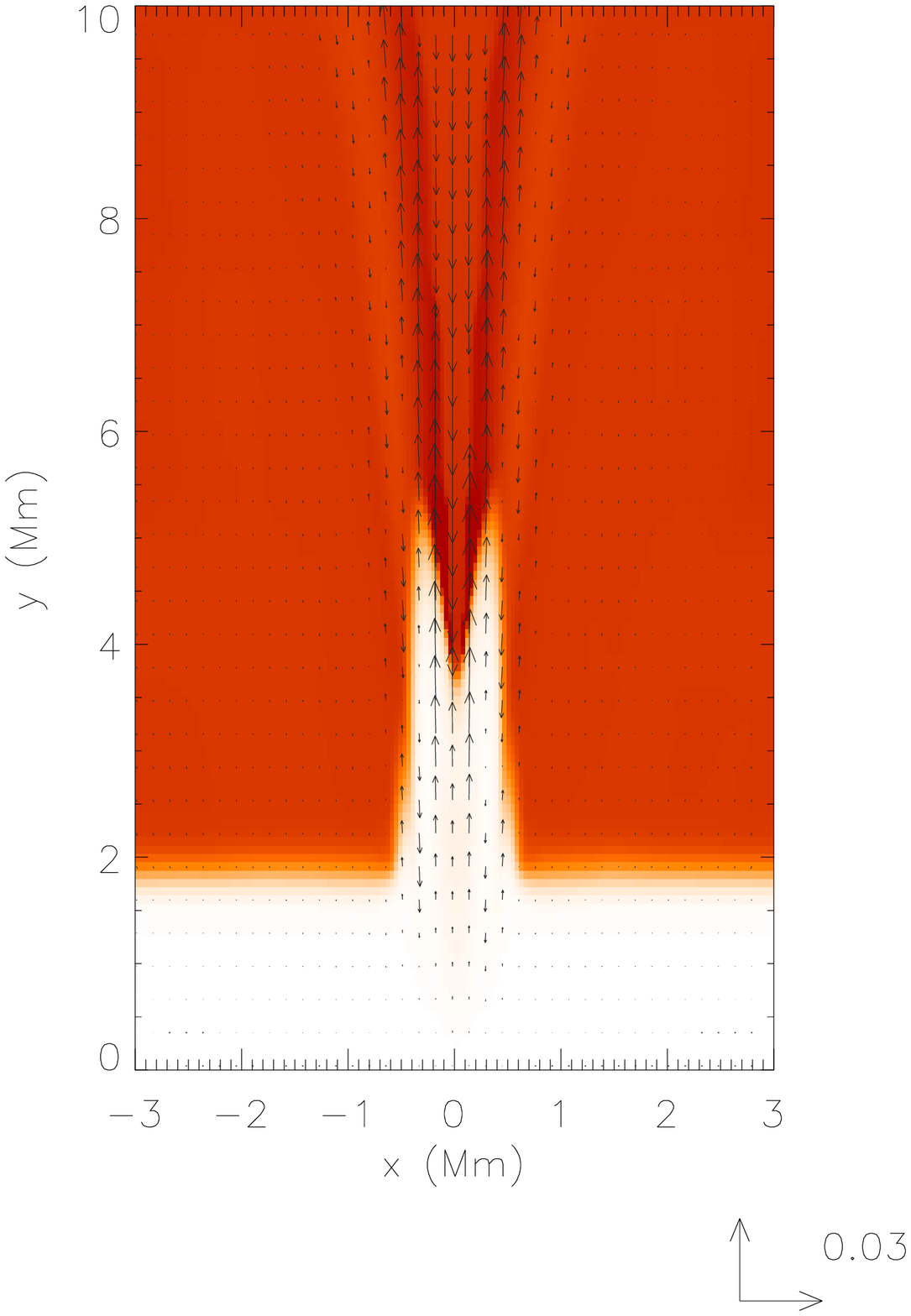}
\includegraphics[width=6.1cm,height=7.2cm]{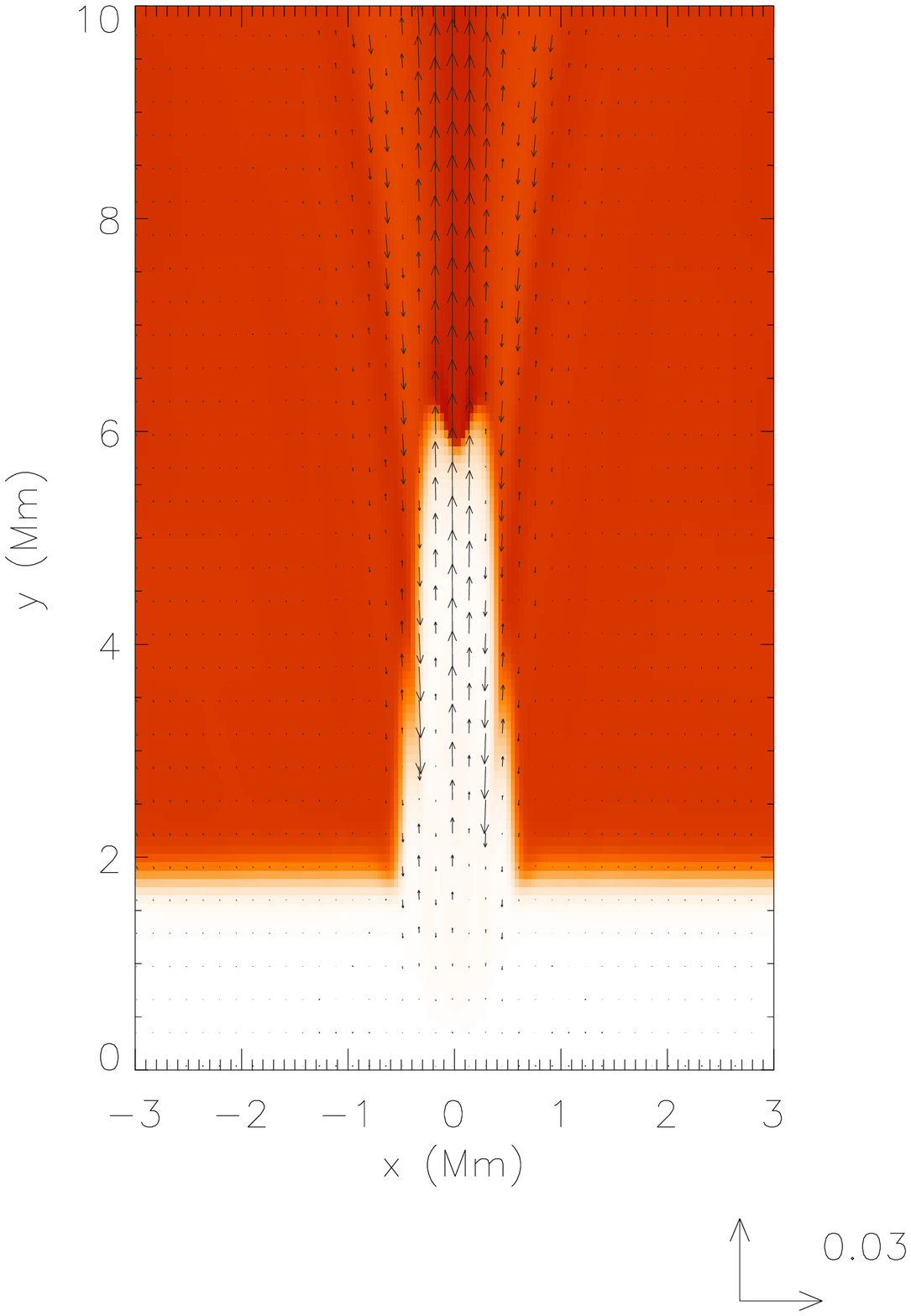}
} \caption{
\small
Temperature (colour map plot) and velocity (arrows) profiles at
$t=50$ s, $t=190$ s, $t=290$ s, $t=360$ s, $t=400$ s, $t=500$ s,
$t=790$ s, and $t=890$ s for $(x_{\rm 0}=0, y_{\rm 0}=0.5)$ Mm
and $A_{\rm v} = 30$ km s$^{-1}$.
Temperature is drawn in units of $1$ MK.
The arrow below each panel represents the length of the velocity vector, expressed in units of $1$ Mm s$^{-1}$.
The colour bar is common to all panels.
}
\label{fig:spicule_prof_cent}
\end{figure*}

While observed in the H$_\alpha$ line,
type I spicules can reach up to $4000-12000$ km in height from the solar limb with a mean value of $7200$ $\pm$ $2000$ km
(Pasachoff et al. \cite{pas09}).
On the other hand, the type II spicules dominate in the lower atmospheric layers: they are tallest in coronal holes
reaching a level of 5000 km, while they are observed to be shorter in quiet and active regions of the solar atmosphere
(De Pontieu et al. \cite{dep07b}).

Typical electron temperature and electron density in spicules are
$15000-17000$ K and 2$\cdot 10^{11}$-3.5$\cdot 10^{10}$ cm$^{-3}$, respectively, at altitudes of
$4000-10000$ km above the solar surface (Beckers \cite{bec68}).
As a result, spicules are much cooler and denser than ambient coronal plasma.

Spicules arise at a level of about $2000$ km with an average speed of $25$ km s$^{-1}$, reach maximum heights, and then either fade away
or descend back to the
photosphere with the same speed (Beckers \cite{bec68}, Pasachoff et al. \cite{pas09}). A typical lifetime of type I
spicules is 5-15 min, but some spicules may live for longer or shorter periods. Pasachoff et al. (\cite{pas09}) find that the distribution of
a spicule lifetime exhibits a peak at $\sim$ 7 min.
On the other hand, the type II spicules
are characterized by
a shorter life time, which is about 10-150 s, and they reach higher velocities of $50-100$
km s$^{-1}$ (De Pontieu et al. \cite{dep07b}).

Most spicules show a double thread structure during their evolution. This double structure was reported for the first time
by Tanaka (\cite{Tanaka1974})
and then by Dara et al. (\cite{Dara1998}). Recently, high-resolution observations performed by Hinode confirmed
that most spicules exhibit the double thread structures
(Suematsu et al. \cite{Suematsu2008}). Suematsu et al. (\cite{Suematsu2008}) find that the separation of some of
the double thread spicules vary in time, repeating a single-thread phase and the double-thread one. The width of each thread and
a separation distance between them is a few tenths of arc sec on average.

Another important feature in spicule dynamics is the bi-directional flow. This phenomenon has been found to be typical of mottles
(Tsiropoula et al. \cite{Tsiropoula1994}, Tziotziou et al. \cite{Tziotziou2003,Tziotziou2004}),
the base portions showing downward motions while the tops of the structures have alternating upward and downward motions.
On the other hand, Pasachoff et al. (\cite{pas09}) detected regular, oppositely directed motions at lower levels in
40 limb spicules through high-resolution observations on SST. Therefore, bi-directional motions seem to be typical for spicules.
\section{A numerical model}\label{sect:num_model}
%
%
%
Our model system is taken to be composed of a
gravitationally-stratified solar atmosphere in
a
2D
space $x$-$y$. We restrict ourselves to the ideal
MHD equations:
\beqa
\label{eq:MHD_rho}
{{\partial \varrho}\over {\partial t}}+\nabla \cdot (\varrho{\bf V})=0\, ,
\\
\label{eq:MHD_V}
\varrho{{\partial {\bf V}}\over {\partial t}}+ \varrho\left ({\bf V}\cdot \nabla\right ){\bf V} =
-\nabla p+ \frac{1}{\mu}(\nabla\times{\bf B})\times{\bf B} +\varrho{\bf g}\, ,
\\
\label{eq:MHD_B}
{{\partial {\bf B}}\over {\partial t}}= \nabla \times ({\bf V}\times{\bf B}),\,\,\,\nabla\cdot{\bf B} = 0\, ,
\\
\label{eq:MHD_p}
{\partial p\over \partial t} + \nabla\cdot (p{\bf V}) = (1-\gamma)p \nabla \cdot {\bf V}\, ,
\\
\label{eq:MHD_CLAP}
p = \frac{k_{\rm B}}{m} \varrho T\, .
\eeqa
Here ${\varrho}$ is mass density, ${\bf V}$ is flow velocity,
${\bf B}$ is the magnetic field, $p$ is gas pressure, $\gamma=5/3$ is the
adiabatic index, ${\bf g}=(0,-g)$ is gravitational acceleration of
its value $g=2.72\cdot 10^2$ m s$^{-2}$, $T$ is temperature, $m$ is
mean particle mass and $k_{\rm B}$ is the Boltzmann's constant.
\subsection {Initial configuration: equilibrium}
%
%
%
We assume that the solar atmosphere is in static equilibrium (${\bf V}_{\rm e}=0$) with a force-free magnetic field, i.e.,
\begin{equation}
\label{eq:B}
(\nabla\times{\bf B}_{\rm e})\times{\bf B}_{\rm e} = 0\, .
\end{equation}
As a result, at this equilibrium the pressure gradient is balanced by the gravity force
\begin{equation}
\label{eq:p}
-\nabla p_{\rm e} + \varrho_{\rm e} {\bf g} = 0\, .
\end{equation}
Here the subscript $_{\rm e}$ corresponds to equilibrium quantities.

Using the ideal gas law and the $y$-component of hydrostatic
pressure balance indicated by Eq.~(\ref{eq:p}), we express
equilibrium gas pressure and mass density as
\beqa
\label{eq:pres}
p_{\rm e}(y)=p_{\rm 0}~{\rm exp}\left( -\int_{y_{\rm r}}^{y}\frac{dy^{'}}{\Lambda (y^{'})} \right)\, ,
\\
\label{eq:eq_rho}
\varrho_{\rm e} (y)=\frac{p_{\rm e}(y)}{g \Lambda(y)}\, .
\eeqa
Here
\begin{equation}
\Lambda(y) = k_{\rm B} T_{\rm e}(y)/(mg)
\end{equation}
is the pressure scaleheight, and $p_{\rm 0}$ denotes the gas
pressure at the reference level that is chosen at $y_{\rm r}=10$ Mm.
Note that $y=0$ corresponds to the base of the photosphere.

We adopt a smoothed step-function profile for plasma temperature
%
\begin{equation}
T_{\rm e}(y) = \frac{1}{2} T_{\rm c} \left[1 + d_{\rm t} +
(1 - d_{\rm t}){\rm tanh} \left(\frac{y-y_{\rm t}}{y_{\rm w}}\right) \right],
\end{equation}
where $d_{\rm t}=T_{\rm ch}/T_{\rm c}$ with $T_{\rm ch}$ denoting
the chromospheric temperature at its lower part. The symbol
$T_{\rm c}$ corresponds to the temperature of the solar corona that is
separated from the chromosphere by the transition region of its
width $y_{\rm w} = 200$ km located at $y_{\rm t}=2$ Mm above the
solar surface. We assume 
$T_{\rm ch}=15\cdot 10^3$ K and $T_{\rm c}=3$ MK.

For the initial magnetic field, we adopt the magnetic field model
that was originally described by Priest (1982). See also
Wasiljew \& Murawski (2009) for a recent application of the solar atmosphere model
in the context of coronal loop oscillations.
In this model, we assume that
Eq.~(\ref{eq:B}) is satisfied by a current-free magnetic field
$$\nabla \times \vec B_{\rm e}=0$$
such that
$$
\vec B_{\rm e}=\nabla \times (A\hat {\bf z})\, .
$$
Here $A$ denotes the magnetic flux function
\begin{equation}
A(x,y) = B_{\rm 0}{\Lambda}_{\rm B}\cos{(x/{\Lambda}_{\rm B})} {\rm exp}[-(y-y_{\rm r})/{\Lambda}_{\rm B}]\, .
\end{equation}
%
The equilibrium magnetic field components $(B_{\rm ex},B_{\rm ey})$
are then given by
\beqa\label{eq:B_com}
B_{\rm ex} & = & \hspace{2.5mm} B_{\rm 0} \cos{({x}/{\Lambda}_{\rm B})} \times {\rm exp}[-(y-y_{\rm r})/{\Lambda}_{\rm B}]\, ,\\
B_{\rm ey} & = &             -  B_{\rm 0} \sin{({x}/{\Lambda}_{\rm B})}\times {\rm exp}[-(y-y_{\rm r})/{\Lambda}_{\rm B}]\,
\label{eq:B_com_By}
\eeqa
in addition to $B_{\rm ez}=0$. Here, $B_{\rm 0}$ is the magnetic
field at $y=y_{\rm r}$, and the magnetic scale-height is
${\Lambda}_{\rm B}=2L/\pi$.
We choose $L\simeq 30$ Mm, which corresponds to the size of a supergranular cell. This magnetic field
is predominantly vertical at supergranular boundaries
(${x}/{\Lambda}_{\rm B}=n\pi/2$, where $n=\pm 1, \pm 3,...$), while
it reveals a horizontal canopy structure at supergranular centres
(${x}/{\Lambda}_{\rm B}=n\pi$, where $n=0, \pm 2,...$).

Observations show that type I spicules are mainly formed at the supergranular
boundaries, where the magnetic field is predominantly vertical. A magnetic field is more complicated and
is concentrated in tubes
in quiet regions of the solar photosphere, while at higher altitudes, the magnetic tubes merge
and produce large-scale structures (network and canopy).
As we are interested in the dynamics of top layers of the solar atmosphere implementation of
the simple configuration of Eqs.~(\ref{eq:B_com}) \& (\ref{eq:B_com_By}) is justified.

Figure~\ref{fig:initial_profile} (upper panel) illustrates
spatial profiles of equilibrium mass density $\varrho_{\rm e}(y)$ (colour map) and magnetic field lines,
near supergranular boundaries (arrows).
The coordinate $x$=0 corresponds to the central line between neighbouring supergranular cells.
The mass density experiences a sudden fall off with height $y$ at $y\simeq 0.5$ Mm, and the magnetic field diverges
with $y$.
The bottom panel shows the spatial profile of the equilibrium temperature $T_{\rm e}(y)$, which rises abruptly to its coronal value
at the transition region, $y=2$ Mm.
\subsection{Perturbations}
Since we are studying the rebound shock model of spicule excitation,
we initially perturb (at $t=0$) the above equilibrium by a Gaussian pulse in the vertical component
of velocity, viz.,
\beq
V_{\rm y}(x,y,t=0) = A_{\rm v} \exp\left[ -\frac{(x-x_{\rm 0})^2+(y-y_{\rm 0})^2}{w^2} \right]\, .
\eeq
Here $A_{\rm v}$ is the amplitude of the pulse, $(x_{\rm 0},y_{\rm 0})$ is its initial position and
$w$ denotes its width.
\section{Results of numerical simulations}
Equations (\ref{eq:MHD_rho})-(\ref{eq:MHD_CLAP}) are solved numerically using the code FLASH
(Lee {\it et al.} 2009). This code implements a second-order unsplit Godunov solver (e.g., Murawski \cite{mur02})
with various slope
limiters and Riemann solvers, as well as adaptive mesh refinement (AMR).
We use the monotonized central slope limiter and the Roe Riemann solver (e.g., Toro 1999). We set the simulation box as
$(-3,3)\, {\rm Mm} \times (0,20)\, {\rm Mm}$
and impose boundary conditions fixed in time for all plasma quantities
in the $x$- and $y$-directions, while
all plasma quantities remain invariant along the $z$-direction.
In all our studies we use AMR grid with a minimum (maximum) level of
refinement set to $3$ ($6$). The refinement strategy is based on
controlling numerical errors in mass density.
This results in an excellent resolution of steep spatial profiles and
greatly reduces numerical diffusion at
these locations.
The initial (at $t=0$ s) AMR system is displayed in Fig.~\ref{fig:initial_profile} (lower panel).
Every block consists of $64$ identical numerical cells.

We launch the velocity pulse of Eq.~(14) with $y_0=0.5$ Mm (i.e. lower chromosphere) and $w=0.3$ Mm.
The amplitude of the pulse is taken to be strong enough. We suggest that
the pulse originates at the bottom of the photosphere.
Therefore its amplitude significantly grows with height as a result of the rapid decrease in mass density in the photosphere
(Carlsson \& Stein \cite{carlson97}). The width of the photosphere
(500 km) implies about $4$ pressure scaleheights. Therefore, the amplitude of a velocity signal grows by $\sim$ $7$ times because of
the energy flux conservation in the case of vertical propagation. For example, the velocity of a granular cell of
$1.5-2$ km s$^{-1}$ would lead to
a flow amplitude of $\sim$ $10-15$ km s$^{-1}$ at height of $500$ km.
A stronger initial pulse at the photospheric level would obviously result in a larger amplitude flow at the lower chromosphere.
Roberts (\cite{roberts1979}) proposed that the resonant buffeting of anchored magnetic flux tubes by granular cells
may trigger strong upward flows of $4-5$ km s$^{-1}$ at the base of the photosphere.
These flows would result in magnitudes as large as $\sim$ $30-35$ km s$^{-1}$ in the lower chromosphere.
Other types of activity, such as, a magnetic reconnection, may trigger even stronger pulses.
Therefore, we consider $A_{\rm v} > c_{\rm s}$, where $c_{\rm s}$ is the chromospheric sound speed that is of the order of
$10$ km s$^{-1}$.

Figure~\ref{fig:spicule_prof_cent} displays the spatial profiles of vertical velocity (arrows) and plasma temperature
resulting from the initial velocity pulse that was launched at the point ($x_0=0,y_0=0.5$) Mm,
at which the magnetic field is essentially vertical.
The upper left panel corresponds to $t=50$ s. Chromospheric plasma lags behind the shock front, which rises up to $y=3.4$ Mm.
The reason for the material being lifted up is
the rarefaction of the plasma behind the shock front, which leads to low pressure there.
As a result, the pressure gradient force works against gravity and forces
the chromospheric material to penetrate the solar corona.
The chromospheric plasma reaches the level of $y=6.5$ Mm in the next $100$ s. The upper central panel is drawn
for $t=190$ s, which clearly resembles a spicule-like structure with
the chromospheric temperature and mass density. Its width and mean rising speed are $600-700$ km and
$25$ km s$^{-1}$, respectively. These values are close to the corresponding characteristics of type I spicules.
At $t=190$ s, the plasma already flows downwards because it is attracted by gravity.
\begin{figure}[h]
\begin{center}
\includegraphics[scale=0.45]{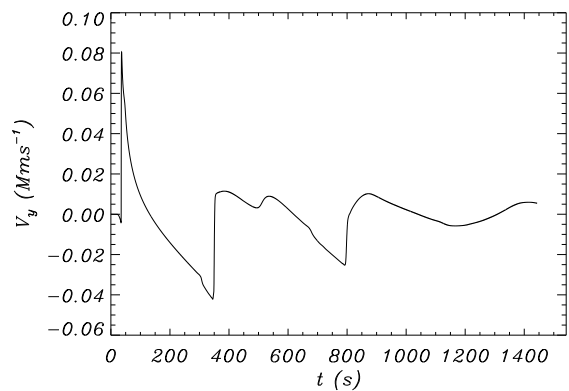}
\caption{\small
Timesignatures of velocity $V_{\rm y}$ collected at ($x=0, y=2.5)$ Mm
for $(x_{\rm 0}=0, y_{\rm 0}=0.5)$ Mm and $A_{\rm v} = 30$ km s$^{-1}$.
Time $t$ and $V_{\rm y}$ are expressed in units of $1$ s and $1$ Mm s$^{-1}$,
respectively.
}
\label{fig:time_profile}
\end{center}
\end{figure}
\begin{figure}[h]
\begin{center}
\includegraphics[scale=0.45]{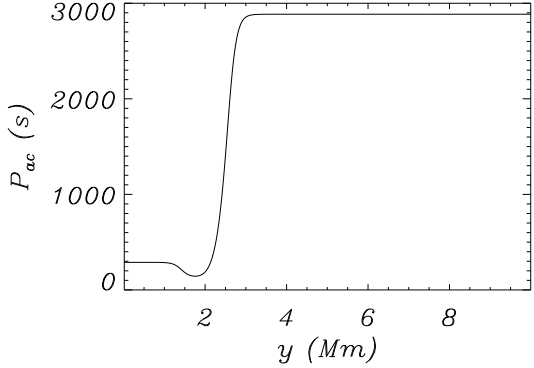}
\caption{\small
Acoustic cutoff wave period $P_{\rm ac}$ (in units of $1$ s) vs height (in units of $1$ Mm) plotted from Eq.~(16).
}
\label{fig:P_ac}
\end{center}
\end{figure}

The next snapshot (upper right panel) was taken at $t=290$ s.
The spicule already subsided to $y=5.2$ Mm and the plasma flows downward near the spicule centre.
However, as the secondary shock lifts up
the chromospheric material (see two small peaks on the left and right sides of central spicule),
there are upward flows at the spicule boundaries.
The secondary shock seems to be wider than the initial one.
Therefore it lifts up the chromospheric material in the regions out of
the initial pulse. The central part is overcome by the falling material of the spicule, therefore it is dominated by downflows.
There are clear bi-directional flows in the spicule: downward at the centre and upward at the boundaries.
This is fully consistent with the observational findings
(Tsiropoula et al. \cite{Tsiropoula1994}, Tziotziou et al. \cite{Tziotziou2003,Tziotziou2004}, Pasachoff et al. \cite{pas09}).
After this moment, the central part of the spicule continues to subside,
while the two small peaks from the previous snapshot become taller and exhibit a clear double structure in the region of
$3.5\,{\rm Mm} <y<4.5$ Mm.
This scenario is consistent with the observational data of Tanaka (\cite{Tanaka1974}),
Dara et al. (\cite{Dara1998}), and Suematsu et al. (\cite{Suematsu2008}).

During the subsequent time intervals, the two peaks rise, merge, and reach a maximum height of $y\simeq 7$ Mm
as seen
at $t=500$ s.
Afterwards the material again falls back at the central part, while the next shock forces the plasma to move upwards
at the spicule boundaries.
We see again two peaks at $t=790$ s in the region of $3.5\,{\rm Mm}<y<5.5$ Mm.
These two peaks rise again and merge to resemble a single spicule at $t=890$ s (lower right panel).
The spicule reveals a clear recurrence-like phenomena with single and double structures and uni-directional and bi-directional flows
occurring and disappearing repeatedly in time until perturbations finally fade away and the whole system relaxes to a static state.

Owing to their large amplitudes, initial pulses steepen rapidly into shocks. Figure~\ref{fig:time_profile} illustrates the $y$-component
of velocity that is collected in time at the detection point $(x=0, y=2.5)$ Mm for the initial pulse amplitude of $A_{\rm v} = 30$ km s$^{-1}$.
The arrival of the shock front at the detection point, $y=2.5$ Mm, is clearly seen at $t\simeq 30$ s.
The second shock front reaches the detection point at $t\simeq 320$ s, i.e., after $\sim$ 5 min
(which fits
the observed $5$-min periodicity in the spicule appearance).
This secondary shock results from the nonlinear wake, which lags behind.
In the linear approximation,
the wake oscillates with the acoustic cut-off frequency
%
\beq\label{eq:Omega_ac}
\Omega_{\rm ac} = {{c_s}\over {2 \Lambda}}\sqrt{1+2{{d \Lambda}\over {dy}}}\, .
\eeq

\begin{figure}[h]
\begin{center}
\includegraphics[scale=0.45]{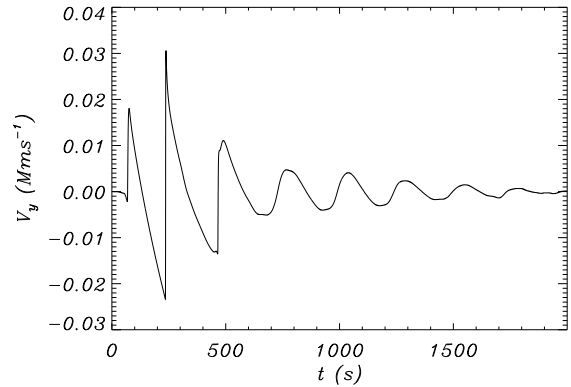}
\caption{\small
Timesignatures of $V_{\rm y}$ collected at ($x=0, y=2.5)$ Mm
for $(x_{\rm 0}=0, y_{\rm 0}=0.5)$ Mm, when $A_{\rm v} = 5$ km s$^{-1}$.
Time $t$ and $V_{\rm y}$ are expressed in units of $1$ s and $1$ Mm s$^{-1}$,
respectively.
}
\label{fig:fft_ts}
\end{center}
\end{figure}

Figure~\ref{fig:P_ac} displays the acoustic cut-off period $P_{\rm ac}= {2\pi/\Omega_{\rm ac}}$ of the model atmosphere vs height.
In the low chromosphere, $P_{\rm ac}\simeq 200$ s and then quickly increases towards the corona.
It is noteworthy that Eq.~(\ref{eq:Omega_ac}) is obtained from the linear analysis, while the nonlinear description may change
the oscillation period of wake. Figure~\ref{fig:fft_ts} shows the $y$-component of velocity that is collected in time
at the detection point $(x=0, y=2.5)$ Mm for the weaker initial pulse with
its amplitude $A_{\rm v} = 5$ km s$^{-1}$. Indeed, the consequent shocks arrive with significantly shorter time intervals
than in the case of $A_{\rm v} = 30$ km s$^{-1}$. The second shock front arrives in $t\simeq 200$ s after the passage of the first one.
The interval between next shocks increases up to 250 s. Thus, the time interval between consecutive shocks depends on
the amplitude of initial pulse, and it is longer for stronger pulses.
\begin{figure*}
{
\includegraphics[width=6.1cm,height=7.2cm]{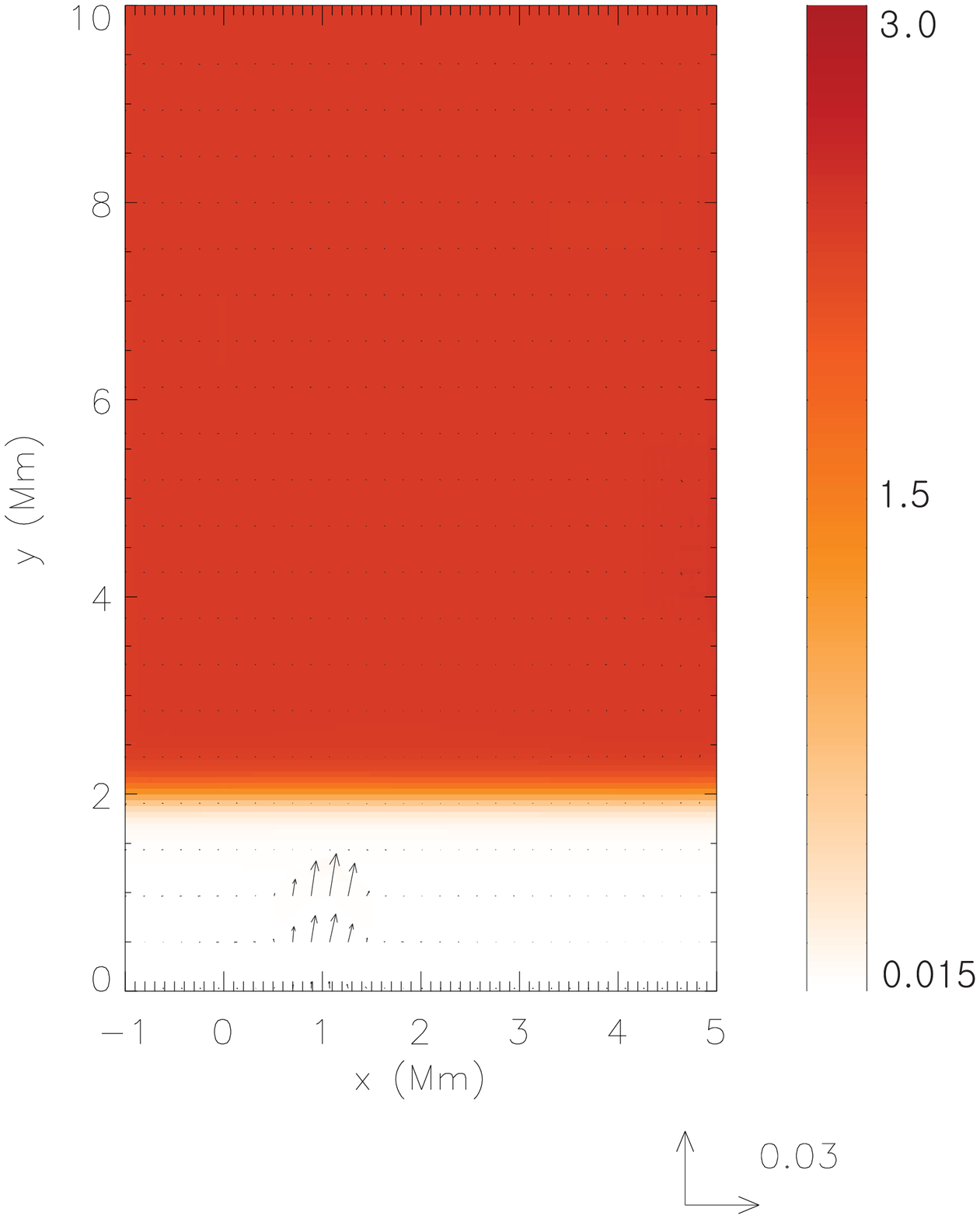}
\includegraphics[width=6.1cm,height=7.2cm]{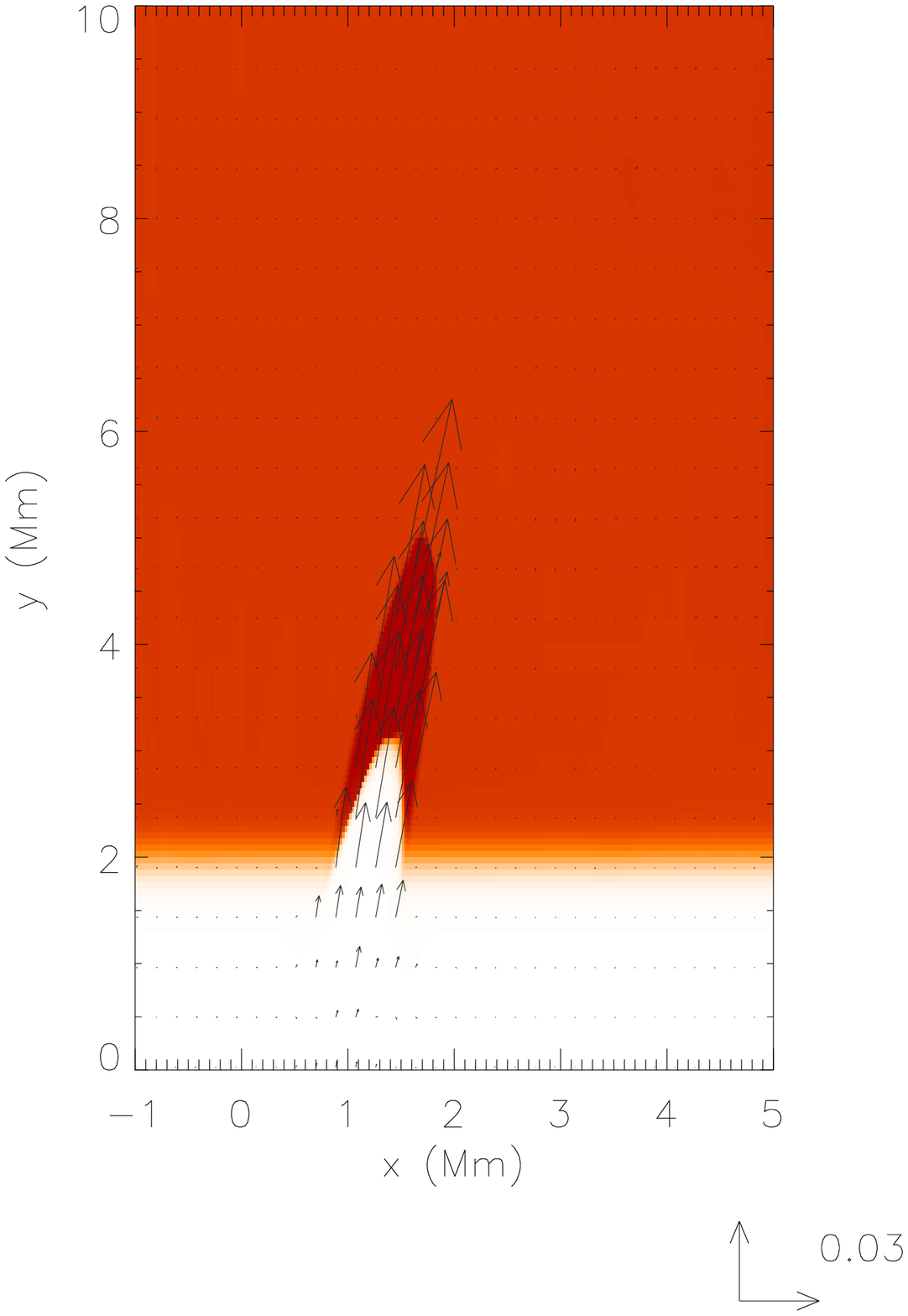}
\includegraphics[width=6.1cm,height=7.2cm]{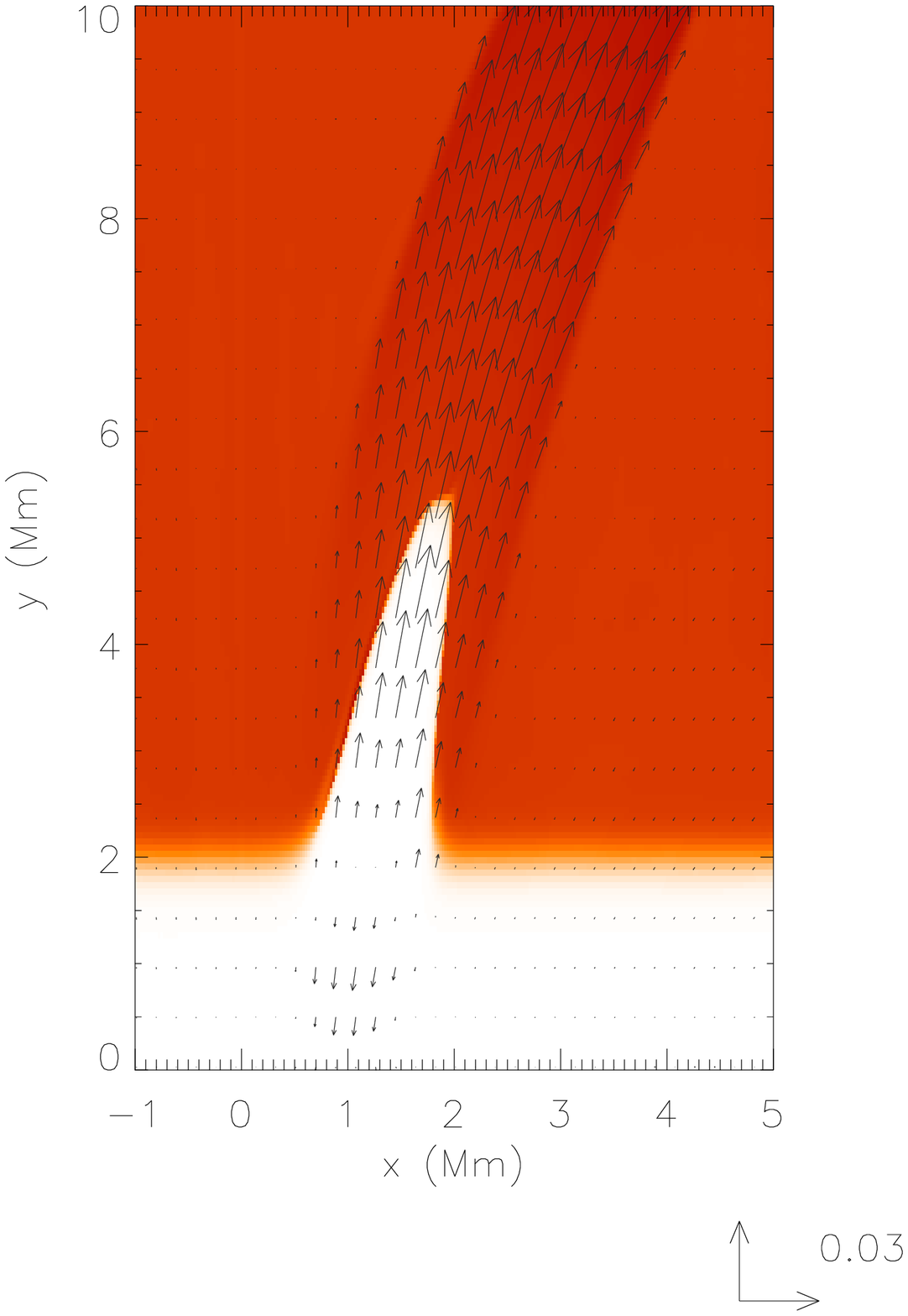}
\includegraphics[width=6.1cm,height=7.2cm]{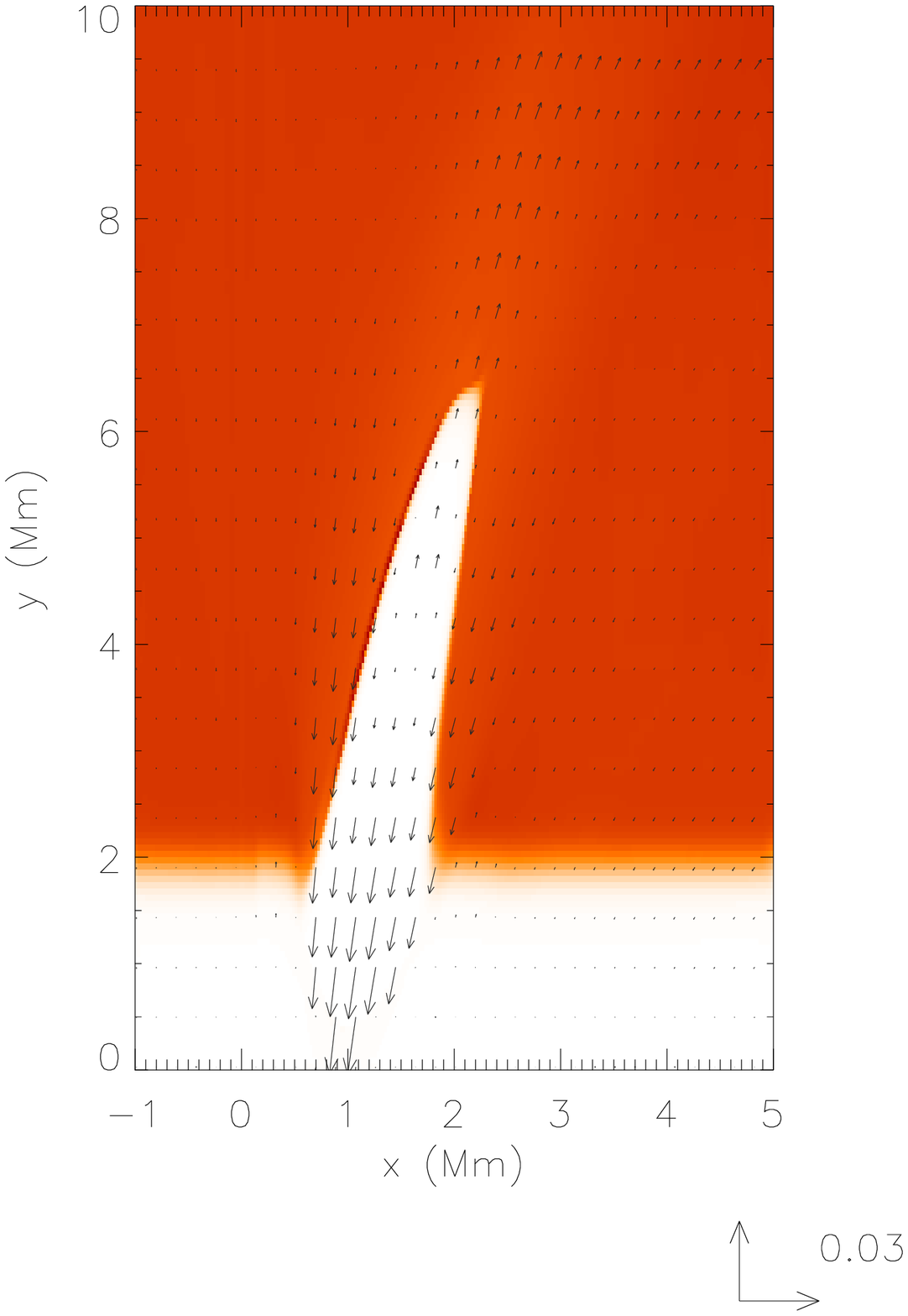}
\includegraphics[width=6.1cm,height=7.2cm]{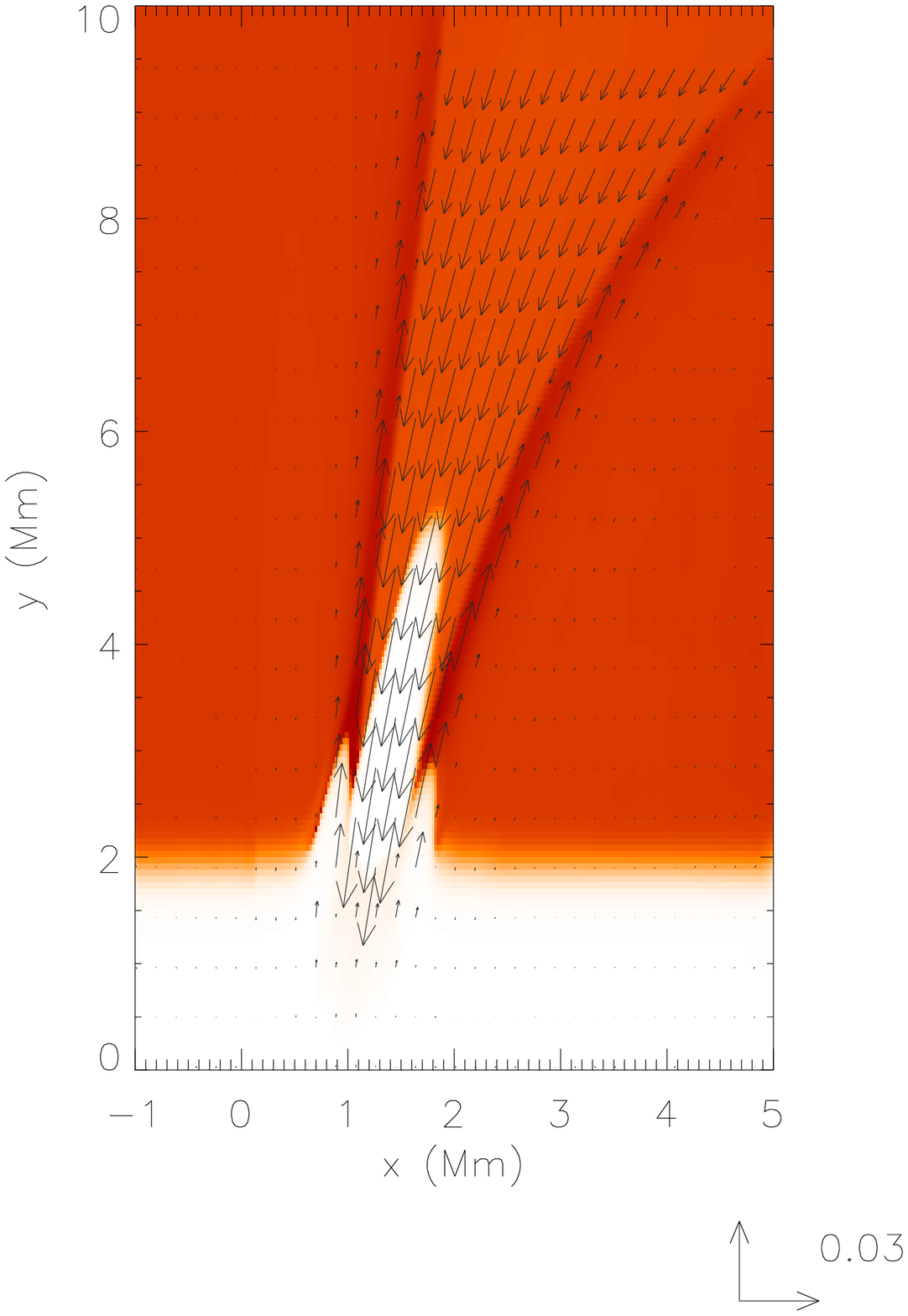}
\includegraphics[width=6.1cm,height=7.2cm]{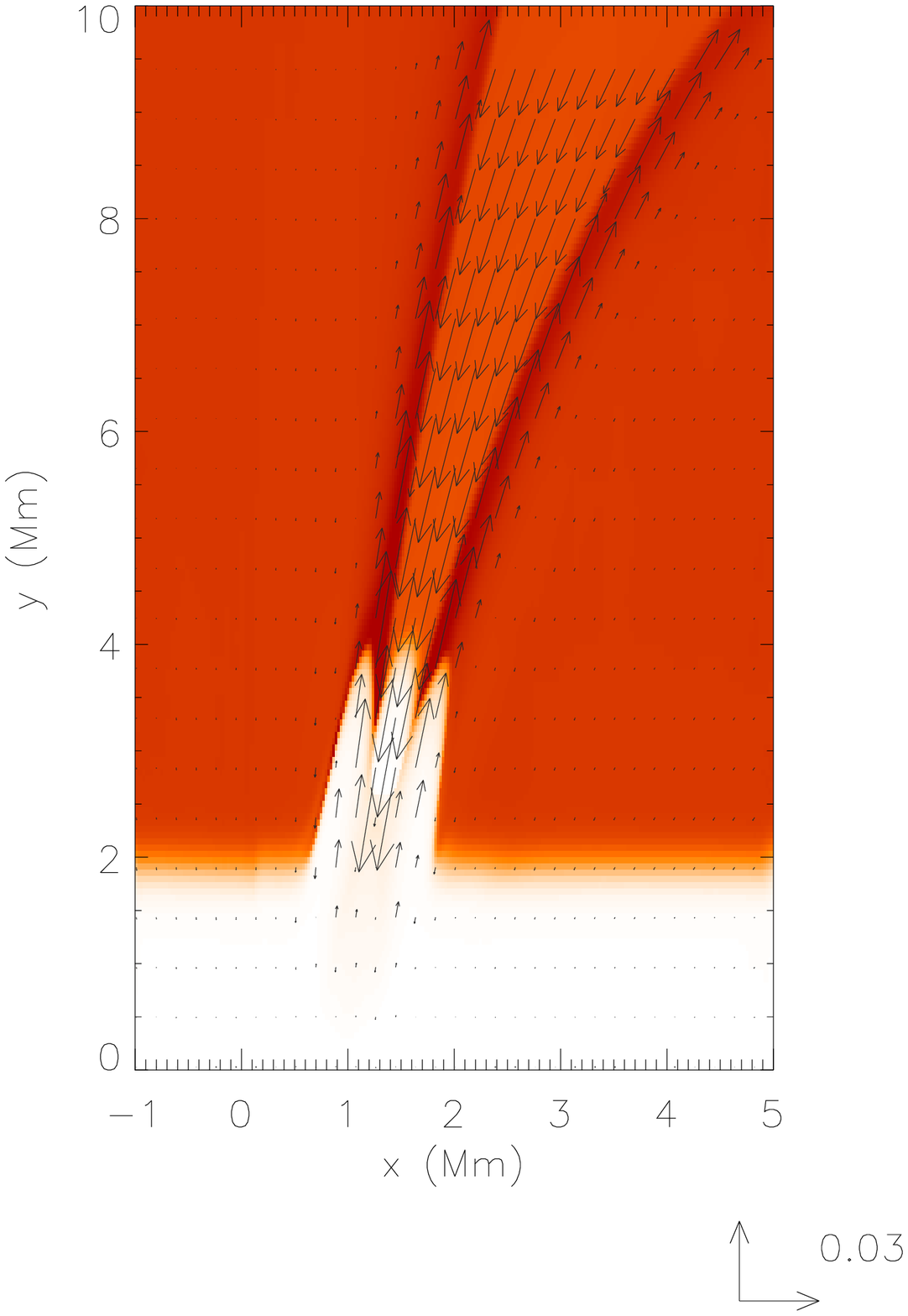}
\includegraphics[width=6.1cm,height=7.2cm]{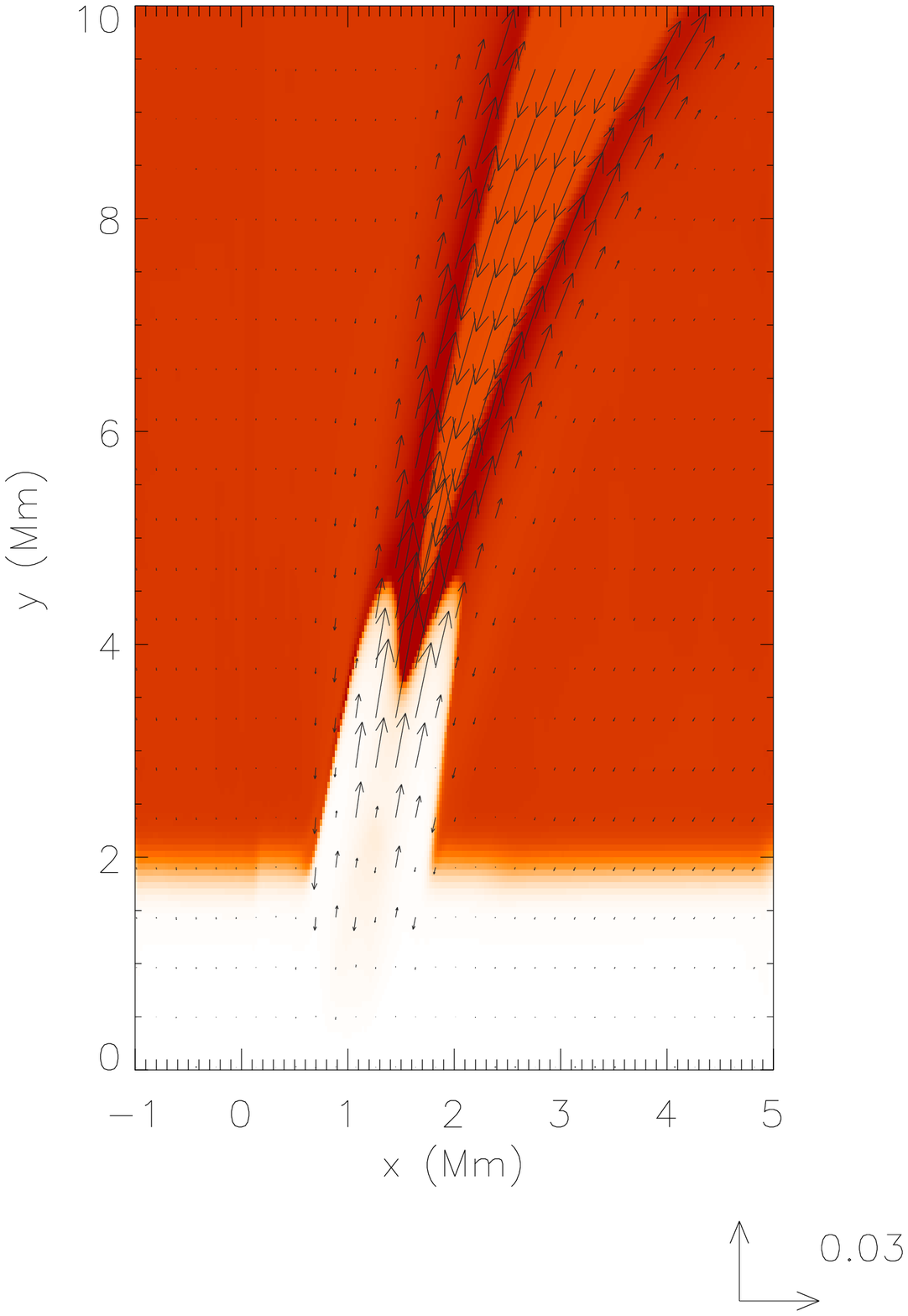}
\includegraphics[width=6.1cm,height=7.2cm]{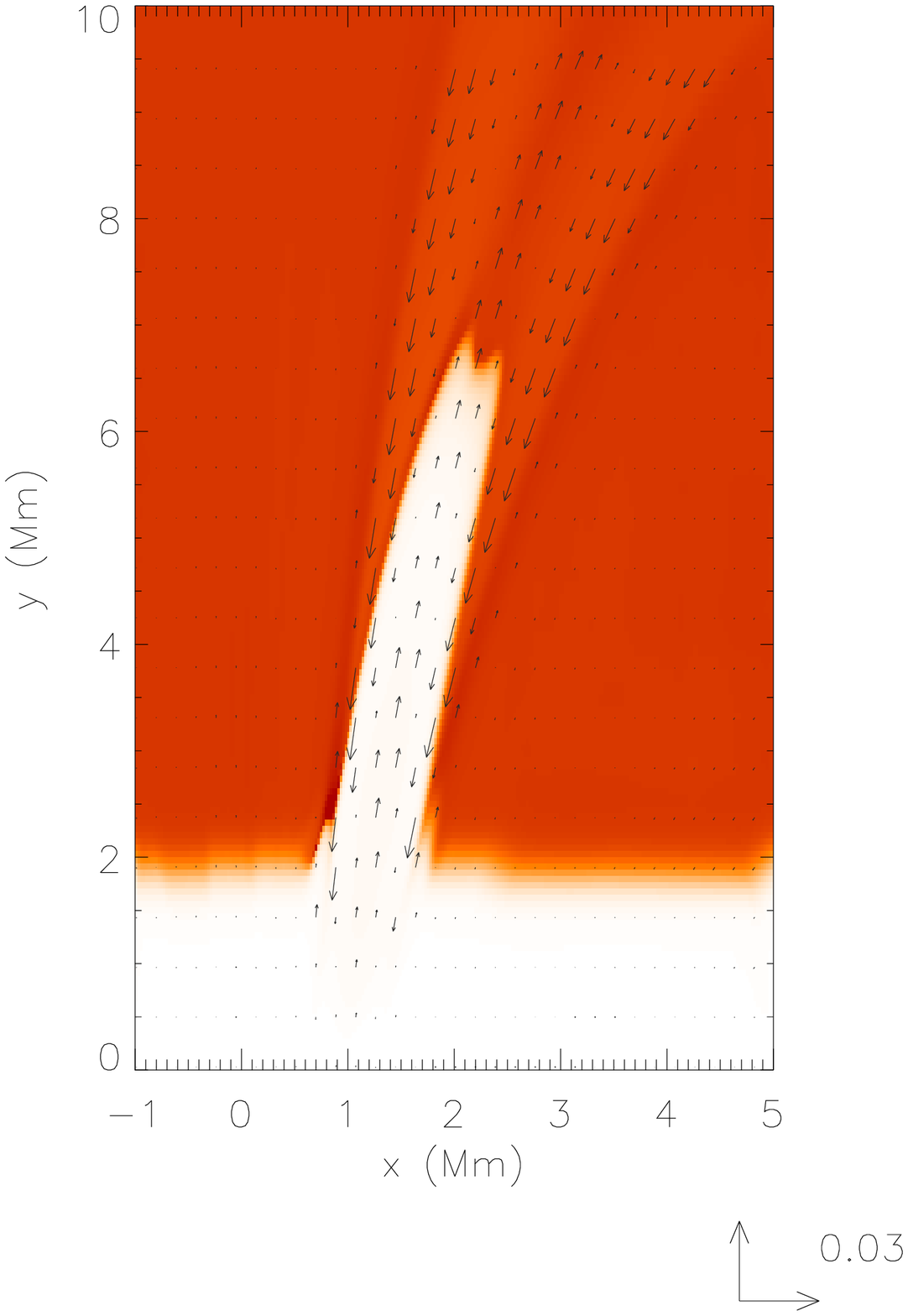}
}
\caption{\small
Temperature (colour map plot) and velocity (arrows) profiles at
$t=20$ s, $t=50$ s, $t=110$ s, $t=180$ s, $t=290$ s, $t=330$ s,
$t=360$ s, and $t=500$ s
for $(x_{\rm 0}=0, y_{\rm 0}=0.5)$ Mm
and $A_{\rm v} = 30$ km s$^{-1}$.
Temperature is drawn in units of $1$ MK.
The arrow below each panel represents the length of the velocity vector, expressed in units of $1$ Mm s$^{-1}$.
The colour bar is common to all panels.
}
\label{fig:spicule_prof}
\end{figure*}

This scenario composes
the building block of the 1D rebound shock model of Hollweg (\cite{hol82}), who proposed
that the secondary shock (or the rebound shock) lifts the transition region higher than the first shock,
thereby resulting in a spicule appearance at observed heights.
The process is well studied in the framework of 1D numerical simulations.
However, our 2D numerical simulations introduce interesting new features in comparison to the 1D rebound shock scenario.

In the case of the simulations whose results are presented
in Fig.~\ref{fig:spicule_prof_cent},
the chromospheric plasma rises vertically, because the pulse was initially launched along
vertical magnetic field lines. However, most spicules are inclined to the vertical.
Actually, magnetic field is vertical only at the centre of a chromospheric network,
while it is inclined in its surroundings. To study what happens around a chromospheric network,
we launched the initial pulse along the inclined magnetic field.
Figure~\ref{fig:spicule_prof} displays the spatial profiles of vertical velocity (arrows) and
plasma temperature at consecutive times after the initial velocity pulse was launched at the point
($x_{\rm 0}=1,y_{\rm 0}=0.5$) Mm, at which the magnetic field is slightly inclined
(see top panel of Fig.~\ref{fig:initial_profile}). The left top panel shows the profiles at $t=20$ s:
the pulse is still propagating within the chromosphere. The next snapshot (central top panel)
is taken at $t=50$ s. The shock already propagates within the corona with high speed,
so it reaches up to the level of $y=6.4$ Mm at this time. On the other hand, the chromospheric material
lags behind the shock along the inclined trajectory and with a much slower speed reaching
the altitude of $y\simeq 3.2$ Mm. The next two snapshots show how the spicule rises to the height
of $\sim$ $6.5$ Mm during the next $130$ s. The spicule is inclined to the vertical, and it follows
the magnetic field structure. The next two snapshots show how the spicule subsides,
while the chromospheric plasma rises at its boundaries due to the secondary shock.
The snapshot at $t=330$ s shows a triple structure
between $y=3.5$ Mm and $y=4$ Mm. The last two snapshots show how the two lateral peaks rise further with height.
First, they form the double structure at $t=360$ s in the region of $3.5$ Mm $<y<4.5$ Mm.
Subsequently, the peaks merge and form a single inclined structure,
which reaches the level of $y=7$ Mm.

\begin{figure}[h]
\begin{center}
\includegraphics[scale=0.45]{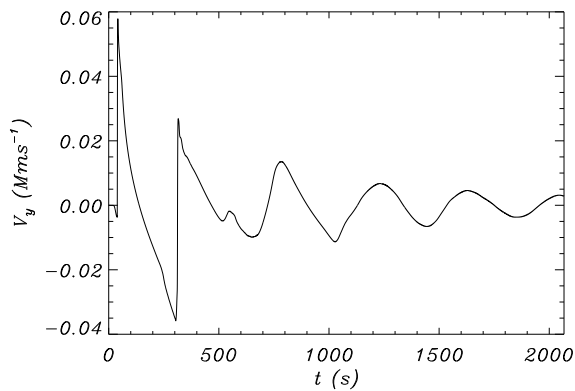}
\caption{\small
Timesignatures of $V_{\rm y}$ collected at ($x=1.15, y=2.5)$ Mm
for $(x_{\rm 0}=1, y_{\rm 0}=0.5)$ Mm
and $A_{\rm v} = 30$ km s$^{-1}$.
Time $t$ and $V_{\rm y}$ are expressed in units of $1$ s and $1$ Mm s$^{-1}$,
respectively.
}
\label{fig:time_profile_off}
\end{center}
\end{figure}

Figure~\ref{fig:time_profile_off} illustrates the $y$-component
of velocity that is collected in time at the detection point $(x=1.15, y=2.5)$ Mm.
This figure is very similar to Fig.~\ref{fig:time_profile}, because both of them reveal that the secondary shock
arrives after $\sim$ $5$ min.
This is because the magnetic field is almost vertical in the chromosphere near supergranular boundaries,
where the initial pulses were launched, while it becomes more inclined in higher coronal regions
(see Fig.~\ref{fig:initial_profile}).
The periodicity of the nonlinear wake, which determines the time interval between consecutive shocks,
is formed in the lower region.
Therefore, we expect that the period of consecutive shocks does not depend significantly on $x$ in
the vicinity of a supergranular boundary.
However, as we already discussed above, the period of the nonlinear wake significantly depends on
the amplitude of the initial pulse.
\section{Discussion}
Our numerical simulations show that the 2D rebound shock model leads to the observed dynamics of type I spicules. The observed rising speed,
width and mean length of spicules can be nicely reproduced with
a sufficiently stronger initial velocity pulse.
The 1D models may also produce the same properties.
However, the advantage of the 2D model is that it may produce other observed properties of spicules,
such as a double (sometimes multi) structure (Tanaka \cite{Tanaka1974}, Dara et al. \cite{Dara1998},
Suematsu et al. \cite{Suematsu2008}) and bi-directional flows (Tsiropoula et al. \cite{Tsiropoula1994},
Tziotziou et al. \cite{Tziotziou2003,Tziotziou2004}, Pasachoff et al. \cite{pas09}).
This is clearly seen in Figs.~\ref{fig:spicule_prof_cent} \& \ref{fig:spicule_prof}.
The multi-structure and bi-directional flows arise when the chromospheric material, which was raised
by the initial shock, falls back due to the gravity, while the secondary shock is lifting up another
portion of chromospheric plasma. The chromospheric plasma is raised up by the pressure gradient behind
the shock. Therefore, the shock propagates with a higher speed that is close to the coronal sound speed,
while the chromospheric plasma is rising with lower velocity, as observed in spicules.

\begin{figure}[h]
\begin{center}
\includegraphics[scale=1.00]{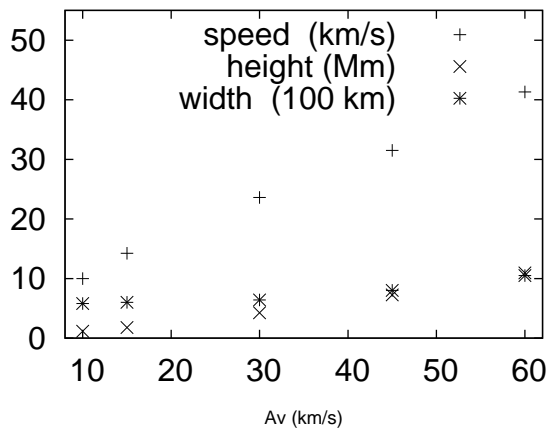}
\caption{
\small
Simulated rising speed (crosses, in units of $1$ km s$^{-1}$), maximum height (x, in $1$ Mm), and
width of spicules (asterisks, in $100$ km)
vs initial amplitude of the velocity pulse $A_{\rm v}$ (in $1$ km s$^{-1}$) for ($x_{\rm 0}=0, y_{\rm 0}=0.5)$ Mm.
Height is measured from the transition region, therefore spicule heights from the solar surface can be produced
by adding of $2$ Mm.
}
\label{fig:spic}
\end{center}
\end{figure}

The mean rising speed and maximum height of spicules significantly depend on a pulse amplitude $A_{\rm v}$.
Figure~\ref{fig:spic} illustrates the dependence of a rising speed (crosses), maximum height measured from
the transition region (x's), and width of spicules (asterisks) on $A_{\rm v}$.
The speeds, height, and width are expressed in units of $1$ km s$^{-1}$, $1$ Mm, and $0.1$ Mm, respectively.
The mean speed and maximum height strongly depend on $A_{\rm v}$ and both grow with $A_{\rm v}$.
On the other hand, the width varies only slightly with $A_{\rm v}$. It turns out that the observed upward speed of
a spicule in the corona ($25$ km s$^{-1}$) and its mean height of $\sim$ 7 Mm
(evaluated from the photospheric level) can
be achieved for $A_{\rm v}=30$ km s$^{-1}$. The initial amplitude of velocity pulse seems to be quite large
for the low chromosphere.
Pulses from granular motions or p-modes can not reach such high amplitudes at $\sim$ 0.5 Mm level;
however, resonant buffeting of granules on anchored magnetic tubes (Roberts \cite{roberts1979}) and/or
magnetic reconnection may produce such strong pulses.

The 2D rebound shock model suggests a quasi-periodic rise of spicules with the period of $\sim$ $5$ min,
which is consistent with observations (Kulidzanishvili \& Nikolsky \cite{kul78}, De Pontieu et al. \cite{dep04},
Xia et al. \cite{xia05}).
In our simulations, this periodicity results from a nonlinear wake that is formed behind
a leading pulse rather than from p-mode leakage.
The period of the wake strongly depends on the amplitude of the initial pulse and is longer for stronger pulses
(see Figs. ~\ref{fig:time_profile} \& \ref{fig:fft_ts}).
The time interval between first and secondary shocks is 200 s and 300 s for the pulses with initial amplitudes
of $5$ km s$^{-1}$ and $30$ km s$^{-1}$, respectively. Thus, the periodicity of spicule appearance should be
$3-5$ min and even longer, depending on a pulse amplitude.
This is consistent with high-resolution observations on SST (Rouppe van der Voort et al. \cite{rouppe2007}).

In our simulations, the initial pulse was launched within the chromosphere, where photospheric magnetic tubes
already merged and produced a larger
scale network magnetic field. The field is nearly vertical at supergranular boundaries and has an almost
horizontal canopy structure above the supergranular cells.
Therefore, our arcade model of magnetic field is a good approximation in this region.
The network structure of the magnetic field is probably
the reason spicules mostly appear near supergranular boundaries; the horizontal magnetic field
prohibits raising chromospheric material, therefore spicules may only appear at the boundaries,
where the magnetic field is predominantly vertical. It would be also interesting to launch pulses at
the photospheric level; however, it requires considering a more complicated magnetic field structure
as it is concentrated in tubes there. This should be the subject of future studies.

We must note here that the model is quite simplified as the equations do not include such terms as plasma partial ionization,
thermal conduction, and radiation. The presence of neutral atoms in plasma is known to change its dynamical and
physical properties significantly (Braginskii 1965, Haerendel \cite{Haerendel1992}, Khodachenko \& Zaitsev 2002,
Khodachenko et al. 2004). Different interaction of electrons, ions, and neutral atoms with the magnetic field and
with each other causes the main specifics of the partially ionized plasma MHD, which differs
significantly from the fully ionized plasma case. However, the effect is more pronounced across the magnetic field,
so it may have little influence on the rebound shock model. On the other hand, radiation and thermal conduction may change the scenario
a lot (Sterling \cite{Sterling1990,Sterling1993}, Cheng \cite{cheng1992}, Heggland et al. \cite{heg07});
therefore, they should be included in future models in order to have a comprehensive description of spicule formation process.
\section{Conclusions}
We performed 2D numerical simulations of velocity pulse propagation in the solar atmosphere.
The strong longitudinal pulse was launched at the lower part of the chromosphere, and its consequent propagation
along transition region into the corona was traced.
The pulse quickly steepens into shock, which influences the dynamics of the upper chromosphere/transition region;
namely, the rarefied tail behind the shock front leads to the pressure gradient above the transition region,
which lifts up the chromospheric material in the form of spicule. The strong initial pulse may raise
the chromospheric plasma up to observed heights ($6-7$ Mm) and with observed speed ($25$ km s$^{-1}$).
The spicule begins to falls back after some time, while the secondary shock (rebound shock) lifts up
another portion of the chromospheric material.
The superposition of falling off and rising plasma portions produces the observed double structure and bi-directional flows in spicules.
The simulated spicule exhibits a temporal development with double and single structures (sometime even triple) that are akin to
the observational data.
The model predicts quasi-periodic raising of spicules with the period of the nonlinear wake that is formed behind
a leading pulse in the stratified atmosphere. The periodicity strongly depends on the initial amplitude of the pulse and can be in
the range of $3-5$ min and even longer. We believe that future sophisticated numerical simulations may give a more complete picture of
spicule formation in the frame of the 2D and 3D rebound shock models.

\begin{acknowledgements}
We are grateful to an anonymous referee and Dr. Hardi Peter whose comments helped us to improve the paper.
The authors express their gratitude to Prof. Serge Koutchmy for useful suggestions.
The work of KM was supported by the Polish Ministry of Science
(the grant for years 2007-2010). The work of TZ was supported by the Austrian Fond zur F\"orderung
der Wissenschaftlichen Forschung (project P21197-N16) and the Georgian National Science
Foundation grant GNSF/ST09/4-310.
\end{acknowledgements}
\end{document}